\documentclass[pdflatex,sn-mathphys-num]{sn-jnl}

\usepackage{graphicx}%
\usepackage{multirow}%
\usepackage{amsmath,amssymb,amsfonts}%
\usepackage{amsthm}%
\usepackage{mathrsfs}%
\usepackage[title]{appendix}%
\usepackage{xcolor}%
\usepackage{textcomp}%
\usepackage{manyfoot}%
\usepackage{booktabs}%
\usepackage{algorithm}%
\usepackage{algorithmicx}%
\usepackage{algpseudocode}%
\usepackage{listings}%
\usepackage{makecell}

\theoremstyle{thmstyleone}%
\theoremstyle{thmstyletwo}%

\theoremstyle{thmstylethree}%

\begin{document}

\title[Article Title]{ Vacancy-induced suppression of CDW order and its impact on magnetic order in kagome antiferromagnet FeGe}


\author[1]{\fnm{Mason L.} \sur{Klemm}}
\equalcont{These authors contributed equally to this work.}
\author[2]{\fnm{Saif} \sur{Siddique}}
\equalcont{These authors contributed equally to this work.}
\author[3]{\fnm{Yuan-Chun} \sur{Chang}}
\author[1]{\fnm{Sijie} \sur{Xu}}
\author[1]{\fnm{Yaofeng} \sur{Xie}}
\author[1]{\fnm{Tanner} \sur{Legvold}}
\author[2]{\fnm{Mehrdad T.} \sur{Kiani}}
\author[4]{\fnm{Feng} \sur{Ye}}
\author[4]{\fnm{Huibo} \sur{Cao}}
\author[4]{\fnm{Yiqing} \sur{Hao}}
\author[4]{\fnm{Wei} \sur{Tian}}
\author[5]{\fnm{Hubertus} \sur{Luetkens}}
\author[4]{\fnm{Masaaki} \sur{Matsuda}}
\author[1]{\fnm{Douglas} \sur{Natelson}}
\author[5]{\fnm{Zurab} \sur{Guguchia}}
\author[3,6]{\fnm{Chien-Lung} \sur{Huang}}
\author[1]{\fnm{Ming} \sur{Yi}}
\author[2,*]{\fnm{Judy J.} \sur{Cha}} 
\author[1,*]{\fnm{Pengcheng} \sur{Dai}}
\affil[1]{\orgdiv{Department of Physics \& Astronomy and Smalley-Curl Institute}, \orgname{Rice University}, \orgaddress{ \city{Houston}, \state{TX}, \country{USA}}}

\affil[2]{\orgdiv{Department of Materials Science and Engineering}, \orgname{Cornell University}, \orgaddress{\city{Ithaca}, \state{NY}, \country{USA}}}
\affil[3]{\orgdiv{Department of Physics and Center for Quantum Frontiers of Research \& Technology (QFort)}, \orgname{National Cheng Kung University}, \city{Tainan}, \country{Taiwan}}
\affil[4]{\orgdiv{Neutron Scattering Division}, \orgname{Oak Ridge National Laboratory}, \city{Oak Ridge}, \state{TN}, \country{USA}}
\affil[5]{Laboratory for Muon Spin Spectroscopy, PSI Center for Neutron and Muon Sciences, Villigen PSI, Switzerland}
\affil[6]{\orgdiv{Taiwan Consortium of Emergent Crystalline Materials}, \orgname{National Science and Technology Council}, \city{Taipei}, \country{Taiwan}}
\affil[*]{e-mail: jc476@cornell.edu, pdai@rice.edu}


\abstract{Two-dimensional (2D) kagome lattice metals are interesting because they display flat electronic bands, Dirac points, Van Hove singularities, and can have interplay between charge density wave (CDW), magnetic order, and superconductivity. In kagome lattice antiferromagnet FeGe, a short-range CDW order was found deep within an antiferromagnetically ordered state, interacting with the magnetic order. Surprisingly,
 post-growth annealing of FeGe at 560$^\circ$C can suppress the CDW order while annealing at 320$^\circ$C induces a long-range CDW order, with the ability to cycle between the states repeatedly by annealing. Here we perform transport, neutron scattering, scanning transmission electron microscopy (STEM), and muon spin rotation ($\mu$SR) experiments to unveil the microscopic mechanism of the annealing process and its impact on magneto-transport, CDW, and magnetic properties of FeGe. We find that 
 560$^\circ$C  annealing creates germanium vacancies uniformly distributed throughout the FeGe kagome lattice, which prevent the formation of Ge-Ge dimers necessary for the CDW order. Upon annealing at 320$^\circ$C, the system segregates into
 stoichiometric FeGe regions with long-range CDW order 
 and regions with stacking faults that act as nucleation sites for the CDW. The presence or absence of CDW order greatly affects the anomalous Hall effect, incommensurate magnetic order, and spin-lattice coupling in FeGe, thus placing FeGe as the only known kagome lattice material with a tunable CDW and magnetic order,  potentially useful for sensing and information transmission. }




\maketitle

\section{Introduction}\label{sec1}
In most correlated electron materials such as hole-doped high-transition temperature copper oxide and iron-based superconductors \cite{keimer_quantum_2015,armitage_progress_2010,dai_antiferromagnetic_2015}, electronic, magnetic, transport, and lattice properties change dramatically as a function of chemical doping, forming a complex electronic phase diagram where intertwined orders of charge-spin-lattice degrees of freedom compete and coexist \cite{fradkin_colloquium_2015}. In electron-doped copper oxide superconductors, however, chemical doping alone is insufficient, and annealing the as-grown sample in a low-oxygen environment, which removes a tiny amount of oxygen ($\sim 1 \%$), is necessary to tune the system from an antiferromagnetic (AFM) metal into a superconductor without magnetic order \cite{tokura_superconducting_1989,takagi_superconductivity_1989}. Annealing 
the electron-doped superconductor in a high-oxygen environment can reverse the effect by adding oxygen back to the system, yielding a non-superconducting antiferromagnet \cite{armitage_progress_2010}. 
While the microscopic origin of such annealing processes remains under debate \cite{tokura_superconducting_1989,takagi_superconductivity_1989,higgins_role_2006,kang_microscopic_2007,guarino_superconductivity_2022}, it is remarkable (and rare) that such a minor modification of the oxygen content in electron-doped copper oxides can so dramatically affect their electronic and magnetic properties \cite{armitage_progress_2010}. Similarly, annealing and quenching van der Waals magnet Fe$_5$GeTe$_2$ is also found to form and destroy vacancy order 
that dramatically affects the electronic structures of the system \cite{wu_reversible_2024}.

Recently, there has been much excitement on kagome lattice antiferromagnet FeGe (the B35 phase) \cite{ohoyama_new_1963} because this correlated electron material exhibits a charge density wave (CDW) order deep within the AFM ordered phase that strongly couples with the magnetic order (Fig. 1a-d) \cite{bernhard_neutron_1984,bernhard_magnetic_1988,teng_discovery_2022,yin_discovery_2022,teng_magnetism_2023,chen_competing_2024,PhysRevX.14.011043,PhysRevLett.132.266505,teng_spin-charge-lattice_2024,PhysRevB.110.L041121}, much different from the usual CDW order observed above or at the magnetic ordering temperature in copper and nickel oxides  \cite{tranquada_cuprate_2020,tranquada_simultaneous_1994,zhang_intertwined_2020}.  
FeGe has a CoSn type structure \cite{meier_flat_2020} in the $P6/mmm$ space group with two Ge positions (Fig. 1a, Ge1 in the Fe$_3$Ge layer and Ge2 between the Fe$_3$Ge layers) \cite{ohoyama_new_1963,bernhard_neutron_1984,bernhard_magnetic_1988}.
With decreasing temperature, the as-grown FeGe exhibits collinear A-type AFM order, {\it i.e.} ferromagnetic order within each Fe$_3$Ge layer that couples antiferromagnetically along the $c$-axis, below the N\'eel temperature $T_N \approx 410$ K (Fig. 1b,c) and then develops incommensurate AFM structure along the $c$-axis below $T_{canting}\approx 60$ K, forming a double-cone AFM structure (Fig. 1d)   \cite{bernhard_neutron_1984,bernhard_magnetic_1988}. It also displays a short-range CDW order below $T_{CDW}\approx 100$ K suggested to be associated with an 
 enhanced Fe moment (Fig. 1c) \cite{teng_discovery_2022} and the formation of Ge-Ge dimers at the Ge1-site along the $c$-axis (Fig. 1a) \cite{miao_signature_2023}. Surprisingly, annealing the as-grown sample in vacuum at 320$^\circ$C changes the short-range CDW into long-range CDW, while annealing at 560$^\circ$C suppresses the CDW order \cite{wu_annealing-tunable_2024,chen_discovery_2024,shi_disordered_2023}. Annealing the CDW-suppressed sample at 320$^\circ$C can re-induce the long-range CDW order \cite{wu_annealing-tunable_2024,chen_discovery_2024}; therefore, the annealing-induced long-range CDW to no CDW transition is reversible much like the annealing-induced AFM order to superconductivity transition in electron-doped copper oxides \cite{higgins_role_2006,kang_microscopic_2007,guarino_superconductivity_2022}. From transport, magnetic susceptibility, heat capacity, X-ray diffraction, and scanning tunneling microscopy measurements, it was argued that large occupational disorders or defects at Ge1-sites may prevent the formation of a long-range CDW order, although a microscopic understanding of the annealing process and its impact on magneto-transport, CDW, and magnetic properties is 
still lacking \cite{wu_annealing-tunable_2024,chen_discovery_2024,shi_disordered_2023}.

Here we perform transport, neutron scattering, scanning transmission electron microscopy (STEM), and muon spin rotation ($\mu$SR) experiments to study the annealing effect on as-grown FeGe.
We find that annealing FeGe at 560$^\circ$C leads to Ge loss that creates point defects (vacancies) on Ge1-sites throughout the sample, preventing the formation of Ge1-Ge1 dimers and CDW order (Fig. 1a,j,m) \cite{miao_signature_2023}. Such a process also dramatically reduces the ordering temperature of the incommensurate AFM structure to $T_{canting}\approx 25$ K from 60 K in as-grown sample (Fig. 1f) \cite{teng_discovery_2022}. For 320$^\circ$C annealed FeGe with long-range CDW order \cite{wu_annealing-tunable_2024,chen_discovery_2024}, the vacancies precipitate and form dislocations and stacking faults (Fig. 1i, 3c,l, 4a,e), thus allowing the majority of the sample to be stoichiometric and the formation of Ge1-Ge1 dimers and long-range CDW order (Fig. 1k-l, 3, 4). The long-range CDW order in FeGe enhances the spin-lattice coupling at the CDW ordering temperature. The magnetic ordered moment associated with incommensurate AFM peaks below 
$T_{canting}$ increases (Fig. 2m-o), and cannot be accounted for by 
the double-cone magnetic structure \cite{bernhard_neutron_1984,bernhard_magnetic_1988}. Furthermore, we observe a massive enhancement of the anomalous Hall effect (AHE) compared with as-grown FeGe \cite{teng_discovery_2022}, but at a temperature below $T_{canting}$ (Fig. 1g), which suggests that this results from topological Hall effect (THE)  \cite{smejkal_anomalous_2022}. These results unveil the microscopic origin of the annealing process and establish FeGe as the only known kagome lattice material with tunable CDW and magnetic order that can dramatically affect the magneto-transport (AHE/THE) properties, thus potentially useful for in-situ sensing and information transmission.

\section{Experimental Results}\label{sec2}
{\bf Transport and magnetic susceptibility measurements.}
We characterize the macroscopic effects of post-growth annealing in FeGe via neutron scattering, transport, and magnetic susceptibility measurements. Four annealing conditions are studied in this work, and are labeled as follows: long-range CDW refers to samples annealed at 320\textdegree C for 96 hours; short-range CDW refers to samples annealed at 320\textdegree C for 8 hours; no-CDW refers to samples annealed at 560\textdegree C for 96 hours, and as-grown samples receive no post-growth annealing. Consistent with previous X-ray diffraction work \cite{wu_annealing-tunable_2024,chen_discovery_2024,shi_disordered_2023},
we find that $T_{CDW}$ is enhanced in long-range CDW samples. Figure 1e compares 
the order parameter scans on the strong CDW Bragg peak $\pmb{Q}=(3.5,0,1.5)$ for long-range and short-range CDW samples, revealing a roughly 15 K increase in $T_{CDW}$ for the long-range CDW ordered sample. A dramatic suppression of the incommensurate AFM order occurs for the no-CDW sample, suggesting a strong interplay between incommensurate magnetic and CDW orders. The order parameter scans at the incommensurate AFM  wavevector $\pmb{Q}=(0,0,0.54)$  show an onset of AFM order in long-range CDW samples at $T_{canting}\approx65$ K while the onset of AFM order occurs at $T_{canting}\approx27$ K in the no-CDW samples (Fig. 1f). However, the A-type AFM order is not much affected by the presence or absence of CDW order (Fig. 2f).

In previous work on as-grown FeGe, an AHE was observed at temperatures approximately below $T_{CDW}$ in the spin-flop phase of the A-type order for a $c$-axis aligned magnetic field \cite{teng_discovery_2022}, thus suggesting that
the AHE in FeGe may be associated with the CDW order similar to that of the $A$V$_3$Sb$_5$ 
CDW superconductors \cite{wilson_avsb_2024,yang_giant_2020,yu_concurrence_2021}.
Our magneto-transport measurements in the spin-flop phase of long-range CDW-ordered FeGe reveal that the onset of AHE occurs around $T_{canting}$ and well below $T_{CDW}$ (Fig. 1g).
Furthermore, the long-range CDW order enhances the AHE by order of magnitude at base temperature ($T$ = 2 K) compared to the as-grown FeGe case (Fig. 1g), reaching a value comparable to that of 
$A$V$_3$Sb$_5$ \cite{yang_giant_2020,yu_concurrence_2021}. 
Given that the AHE occurs around 
 $T_{canting}$, it must arise from conduction electrons scattering off spin fluctuations in the geometrically-frustrated antiferromagnets, as seen in the kagome lattice antiferromagnet YMn$_6$Sn$_6$ \cite{wang_field-induced_2021,ghimire_competing_2020,dally_chiral_2021} and Gd$_2$PdSi$_3$ (breathing kagome) \cite{paddison_magnetic_2022}, and be associated 
 with THE \cite{smejkal_anomalous_2022}. Conversely, no-CDW FeGe shows no evidence of AHE (Fig. 1g), consistent with the weak incommensurate AFM order (Fig. 1f).

\begin{table}[h]
\caption{Mass loss through cycling}\label{tab1}%
\begin{tabular}{@{}llllll@{}}
\toprule
Cycle number & (1) 320\textdegree C & (2) 560\textdegree C & (3) 320\textdegree C & (4) 560\textdegree C & (5) 320\textdegree C\\
\midrule
Mass (mg) & \thead{250.38(1) $\rightarrow$ \\250.36(1)}    & \thead{250.08(1) $\rightarrow$ \\249.72(1)}  & \thead{248.02(1) $\rightarrow$ \\248.01(1)}  & \thead{246.30(1) $\rightarrow$ \\244.95(1)} & \thead{242.76(1) $\rightarrow$ \\242.76(1)} \\
Percentage loss (\%) & \thead{-0.008\%}    & \thead{-0.144\%}   & \thead{-0.004\%}  & \thead{-0.548\%}  & \thead{-0.0\%}\\
\botrule
\end{tabular}
\end{table}

To unveil the microscopic origin for the reversibility of long-range and no-CDW ordered phases through 
the annealing process, we carried out a series of annealing at 560\textdegree C and 320\textdegree C, similar to \cite{wu_annealing-tunable_2024}. Between each annealing sequence, the magnetic susceptibility was recorded for the same single crystal (Fig. 1h). Long-range CDW order is still recoverable after at least five annealing cycles. Detailed mass measurements were taken before and after each annealing process (Table 1). We find no mass loss in the samples after annealing at 320\textdegree C for 96 hours within the tolerance of our scale, whereas annealing at 560\textdegree C for 96 hours results in a loss of mass between 0.14\% and 0.55\%. Between cycles, the samples were sonicated to remove any surface contamination; this can cause some mass loss due to brittle samples breaking into tiny pieces. Only larger single crystals were used in each cycle, which accounts for the different starting mass between each cycle. The observed mass loss is attributed to the removal of a tiny amount of germanium by creating Ge vacancies in FeGe since germanium has a higher vapor pressure than Fe. The Ge vacancies coalesce to form extended defects in samples annealed at 320\textdegree C, while the vacancies are uniformly distributed in samples annealed at 560\textdegree C, indicated by red regions in Fig. 1i-j. We have performed the crystal structure refinements for each annealing condition using single-crystal neutron diffraction data. Consistent with earlier X-ray diffraction work \cite{chen_discovery_2024,shi_disordered_2023}, we find no evidence of the cubic B20 phase of FeGe as an impurity phase. However, Ge ions at the Ge1-site form Ge1-Ge1 dimers below 
the CDW transition temperature as illustrated in Fig. 1k and 1l.

{\bf Unpolarized and polarized neutron scattering experiments.}
2D reciprocal space maps from neutron scattering experiments at 6 K are presented in Fig. 2a-c for long-range, short-range, and no-CDW samples respectively. AFM Bragg peaks are indexed at $\pmb{Q}=(H,0,L+0.5)$. CDW peaks appear at $\pmb{Q}=(H+0.5,0,L), (H+0.5,0,L+0.5), (0,K+0.5,L),$ and $(0,K+0.5,L+0.5)$, where $H, K, L=0, \pm 1, \pm 2, \cdots$ \cite{teng_discovery_2022,miao_signature_2023}. Incommensurate peaks arising from the canted AFM structure appear at $\pmb{Q}=(H,0,L+0.5 \pm \delta)$ where $\delta =0.04$ \cite{bernhard_neutron_1984,bernhard_magnetic_1988,teng_discovery_2022,chen_competing_2024}. 

CDW peaks at $H=0.5$ vanish for the no-CDW sample (Fig. 2c) and the incommensurate peak intensity from the canted AFM structure is greatly diminished compared to the long-range and short-range CDW samples (Fig. 2o). Integrated intensity $\pmb{Q}$-cuts of the strong CDW Bragg peak $\pmb{Q} = (3.5,0,1.5)$ become resolution limited in long-range CDW samples with a Gaussian line shape, whereas as-grown samples possess a Lorentzian line shape with an associated correlation length of 33.2 \r{A} in the plane and 48.7 \r{A} along the $c$-axis from \cite{teng_discovery_2022} (Fig. 2d). The lower bounds on the correlation lengths of long-range CDW samples are $78.2 \pm 0.7$ \r{A} in the plane and $162 \pm 0.9$ \r{A} along the $c$-axis, respectively. The intensities of the long-range CDW and as-grown samples are normalized to one another in Fig. 2d to demonstrate the contrast in line shape. 

To compare the intensity of CDW peaks between long-range, short-range, and no-CDW samples, $\pmb{Q}$-cuts of CDW peaks are normalized with respect to the nuclear Bragg peak (1,0,0) (Fig. 2e). The short-range CDW signal shows more than a 50\% reduction in intensity compared to the long-range CDW signal and the no-CDW sample shows no signal above the background. Order parameter scans of the A-type AFM order using unpolarized neutrons for long-range CDW and no-CDW samples show the magnetic Bragg peaks retain identical intensity above $T_{CDW}$, with only the long-range CDW sample seeing a jump below $T_{CDW}$ (Fig. 2f). In the initial report of CDW in as-grown FeGe, an enhancement of the magnetic moment size was reported below $T_{CDW}$ via unpolarized neutrons \cite{teng_discovery_2022}. Since unpolarized neutrons are unable to separate magnetic contributions from nuclear structure contributions due to the CDW order at the same wave vector \cite{teng_discovery_2022,miao_signature_2023}, we performed neutron polarization analysis to separate the magnetic
and structural contributions of the scattering similar to our prior work on iron pnictides \cite{liu_-plane_2020,lipscombe_anisotropic_2010}. Figure 2g shows the scattering geometry within the $[H,0,L]$ scattering plane for a polarized neutron configuration. The incident neutrons are polarized along the $\pmb{Q} (x)$, perpendicular to $\pmb{Q}$ but in the scattering plane $(y)$, and perpendicular to the scattering plane $(z)$. In this setup, the neutron spin-flip (SF, $\sigma_{x}^{SF}$, $\sigma_{y}^{SF}$, $\sigma_{z}^{SF}$) and non-spin-flip (NSF, $\sigma_{x}^{NSF}$, $\sigma_{y}^{NSF}$, $\sigma_{z}^{NSF}$) scattering cross sections are related to the magnetic scattering along the $y$ ($M_{y}$) and $z$ ($M_z$) directions and nuclear scattering $N$ via
\begin{gather}
\begin{pmatrix}
\sigma _{x}^{SF} - b_1 \\
\sigma _{y}^{SF} - b_1 \\
\sigma _{z}^{SF} - b_1 \\
\sigma _{x}^{NSF} - b_2 \\
\sigma _{y}^{NSF} - b_2 \\
\sigma _{z}^{NSF} - b_2 
\end{pmatrix}
= \frac{1}{R+1}
\begin{pmatrix}
    R & R & 1 \\
    1 & R & 1 \\
    R & 1 & 1 \\
    1 & 1 & R \\
    R & 1 & R \\
    1 & R & R
\end{pmatrix}
\begin{pmatrix}
    M_y \\
    M_z \\
    N
\end{pmatrix}
,
\end{gather}

where $R$ is the quality of the neutron beam polarization (flipping ratio, $R=\sigma _{Bragg}^{NSF}/\sigma _{Bragg}^{SF} \approx 13$), and $b_1$ and $b_2$ are background (and nuclear-spin incoherent and other nonmagnetic background scattering) for SF and NSF channels, respectively \cite{dai_antiferromagnetic_2015}. If the Fe-ordered moments have in-plane components along and perpendicular to the $[H,0,0]$ direction defined as $M_{b \perp}$ and $M_{b||}$, we have $M_y \propto \sin ^2 (\theta) M_{b \perp} + \cos ^2 (\theta) M_c$ and $M_z = M_{b||}$ where $\theta$ is the angle between $\pmb{Q}$ and the $(H,0,0)$ direction with $b_{\perp}$ and $b_{||}$ referring to directions perpendicular and parallel to real space $b$-axis respectively \cite{liu_-plane_2020,lipscombe_anisotropic_2010}. 

Full polarization analysis of the magnetic Bragg peak $\pmb{Q}=(2,0,0.5)$ is shown in Fig. 2h. The direction of the moment in a sample must have some perpendicular component to the neutron momentum transfer direction $\pmb{Q}$ in order to observe SF scattering, i.e. purely magnetic signal. For $\pmb{Q}=(2,0,0.5)$, neutrons polarized along the $``y"$ direction point almost directly along the $c$-axis. The $\sigma_{y}^{SF}$ signal is the same magnitude as the NSF signal, effectively registering as background. The identical magnitude of the $\sigma_{x}^{SF}$ and $\sigma_{z}^{SF}$ means the moment direction is perpendicular to both $x$ and $z$, confirming the sample as an A-type antiferromagnet. 

Figure 2i shows order parameter scans of the AFM peak (2,0,0.5) for as-grown and long-range CDW samples in the SF channel. Both samples demonstrate a similar jump in intensity below $T_{CDW}$ that can be attributed strictly to an increasing magnetic moment, not additional structural scattering from the CDW order since the SF channel only observes magnetic contributions. Fig. 2f compares the AFM order parameter of a long-range CDW and no-CDW sample normalized to the (1,0,0) nuclear Bragg peak using unpolarized neutrons.  Polarization analysis of the incommensurate peaks around [0,0,$0.5\pm \delta$] reveals identical spin-flip scattering intensity in the $y$ and $z$ channels for the long-range CDW sample (Fig. 2j), suggesting no preferential orientation of the moment within the plane, i.e. the in-plane spin component takes on a screw-like character (Fig. 1d). Below $T_{CDW}$, incommensurate magnetic peaks at $\pmb{Q}=(H,0,L+0.5 \pm \delta)$ observed below $T_{canting}$ depend greatly on the post-growth annealing conditions (Fig. 2o).

Figure 2k-l shows $\pmb{Q}$-cuts of the incommensurate Bragg peaks above and below $T_{canting}$ for long-range and no-CDW samples respectively. The order parameter is extracted from these plots in Fig. 1f. Although the double-cone AFM structure can reasonably well describe the incommensurate peaks with a canting angle $\alpha=$ 28\textdegree \ for the long-range CDW sample and $\alpha=$ 16\textdegree \ for the as-grown sample (Fig. 1d, 2n), we did not observe the expected large intensity reduction at
the commensurate AFM $\pmb{Q}=(H,0,L+0.5)$ expected for such a model (Fig. 2m) \cite{bernhard_neutron_1984,bernhard_magnetic_1988,chen_competing_2024}. From Equation 2 (see Methods), we expect a reduction of the commensurate AFM Bragg peak intensities of roughly 22\% for a canting angle of 28\textdegree. However, we observe a 10\% reduction on average and no peak experiences a 22\% reduction within error bars. Additionally, we observe a large intensity gain in the incommensurate peaks for the long-range CDW sample (Fig. 2o) that corresponds to an increased in-plane magnetic moment. We therefore conclude that incommensurate AFM peaks cannot arise from the local moment double-cone AFM structure \cite{chen_competing_2024}.

{\bf Scanning Transmission Electron Microscopy (STEM) measurements.}
To understand the microscopic origin of the annealing process, we use STEM for local structural characterization for the short- and long-range CDW samples (Fig. 3). Four FeGe samples were imaged at $\sim$100 K in the $[110]$ zone axis: as-grown sample annealed at 320\textdegree C and quenched ($A$), and then re-annealed at 560\textdegree C and quenched ($AB$), as-grown sample annealed at 560\textdegree C and quenched ($B$), and then re-annealed at 320\textdegree C and quenched ($BA$). Figure 3a shows the heat-treatment history of the four samples. Consistent with previous work \cite{wu_annealing-tunable_2024,chen_discovery_2024,shi_disordered_2023}
and Fig. 2, the 320\textdegree C heat treatments resulted in a long-range CDW in samples $A$ and $BA$, evidenced by the sharp satellite peaks from the CDW in selected area electron diffraction (SAED) patterns in Fig. 3d,m. Correspondingly, samples $B$ and $AB$ that experienced 560\textdegree C as the last annealing step, did not have satellite peaks in their SAED patterns (Fig. 3g,j), indicating an absence of CDW. 

These four FeGe samples were imaged using bright-field TEM (BF-TEM) to reveal a positive correlation between long-range CDW in samples and the presence of extended defects. The BF-TEM images of $AB$ and $B$ samples (no-CDW, Fig. 3f, i) do not show diffraction contrast from any extended defects. On the other hand, samples with long-range CDW, $A$ and $BA$ (Fig. 3c, l), exhibit diffraction contrast features in the BF-TEM images that are consistent with dislocations and stacking faults found in hexagonal crystals \cite{berghezan_transmission_1961}. We note that these extended defects do not change when imaged at 300 K and 100 K (Supplementary Fig. 5). 

Hexagonal systems can host various types of dislocations, some of which are imperfect and create stacking faults \cite{berghezan_transmission_1961}. These systems can exhibit five types of stacking faults – basal, two pyramidal, and two prismatic types \cite{yin_comprehensive_2017}. Figure 4e shows a Fourier-filtered atomic-resolution STEM image of an imperfect dislocation with a Burgers vector perpendicular to the basal plane and with a magnitude of $c/2$. This dislocation creates a prismatic stacking fault, observable as a change in stacking across the dislocation in the atomic-resolution image. Additionally, pyramidal stacking faults are observed, where the defect is inclined with respect to the incoming electron beam, resulting in bright and dark fringes in the BF-TEM images. 

To further investigate the structural differences from the different annealing conditions, we image the four FeGe samples along the $[110]$ zone axis at atomic-resolution using high-angle annular dark-field (HAADF) STEM at $\sim$100 K (Supplementary Fig. 6). From these images, we constructed a mean $2 \times 2$ unit cell projection by averaging intensities of each atomic column in the images (Fig. 3e,h,k,n). Figure 3b (left half) shows the arrangement of Ge (dark blue), Fe (red) and Ge$|$Fe (light blue) atomic columns in FeGe in the same $[110]$ projection. In HAADF-STEM, a $Z$-contrast imaging method, the intensity of atomic columns scales with their atomic number $Z$ (intensity, $I\propto Z^\alpha$, $\alpha=1.6-2$) \cite{howie_image_1979,hartel_conditions_1996}. Hence, atomic columns with heavier atoms appear brighter. Given that the atomic numbers of Fe and Ge are 26 and 32 respectively, Fe columns will have lower intensity than the Ge$|$Fe columns in HAADF-STEM images. This intensity difference is evident in the simulated HAADF-STEM image (Fig. 3b, right half) and the intensity line profile (Fig. 3o, black curve). By measuring the intensities of Fe and Ge$|$Fe atomic columns from the constructed mean unit cells (Fig. 3o,p), we observe that in samples with long-range CDW, these intensities match those of the simulated image. In samples without CDW, the relative intensity of Ge$|$Fe column is lower, indicating a lower net $Z$ for this atomic column compared to samples with long-range CDW. This suggests the presence of Ge vacancies (owing to the higher vapor pressure of Ge) or substitution of Fe in Ge sites in the Ge$|$Fe columns, reducing the net $Z$ and thereby reducing the column intensity in samples without CDW. 

We hypothesize that the absence of CDW in samples with the final annealing temperature of 
560\textdegree C is related to the presence of point defects (either Ge vacancies or Fe substitutions) in Ge$|$Fe columns. Since the Ge atoms in the Ge$|$Fe columns are displaced along the $c$ axis to form Ge dimers in the CDW state \cite{miao_signature_2023}, these point defects likely prevent the Ge displacement, thereby preventing FeGe from entering a CDW state. Our careful mass measurements on samples with different annealing histories reveal a net mass decrease of $\sim 0.5$\%  after each 560\textdegree C annealing cycle, suggesting the formation of Ge vacancies. When annealed at 320\textdegree C, the vacancies precipitate and form dislocations and stacking faults. In this case, there are no point defects in the Ge$|$Fe columns, which allows the Ge atoms to displace and enable the samples to enter the CDW state.

To gain a microstructural understanding of the coupling between the long-range CDW and extended defects, we performed four-dimensional (4D) STEM at $\sim$100 K. In 4D-STEM, a focused electron probe scans the sample, collecting diffraction patterns at each spatial coordinate in the scan. Figure 4a shows a reconstructed real-space image of the $BA$ FeGe sample. Diffraction patterns from two locations (Figure 4b) – one away from extended defects (spot $1$) and one near a stacking fault (spot $2$) – reveal higher-intensity satellite peaks near the the stacking fault, which is also evident in the line profile (red arrow). 

Since the satellite peak intensity reflects the CDW strength, a higher peak intensity near defects indicates that these defects stabilize the CDW compared to defect-free regions of the crystal. To confirm this, we created virtual apertures (see Methods) to construct real-space Bragg and CDW maps. While the Bragg map (Fig. 4c) shows a real-space intensity distribution consistent with BF-TEM where the defects appear darker due to more scattering \cite{cahn_direct_1965}, the real-space CDW map shows the opposite intensity distribution as the CDW is stronger near the extended defects (Fig. 4d). Given that the sample temperature was $\sim$100 K, close to the CDW transition temperature for FeGe, the higher intensity at the defects suggests that these defects act as the nucleation sites for the CDW, similar to recent observations in CDW transition of $1T$-TaS$_2$ \cite{hart_real-space_2024}. 

To confirm that stronger CDW at the extended defects is not due to local compositional variations, we performed STEM energy dispersive x-ray spectroscopy (STEM-EDX) on the $BA$ sample and found no local compositional variations (Supplementary Fig. 7).

We further analyzed our atomic-resolution HAADF-STEM images to investigate displacement of Ge1 atoms in the Fe$_3$Ge Kagome layer in the CDW state. Since the Ge1 atoms either shift up or down along the $c$-axis while the Fe atoms remain undisplaced, the Ge$|$Fe columns should appear elliptical in the $[110]$ zone axis, due to the projection nature of HAADF-STEM images. We fit two-dimensional Gaussians to all the atomic columns in the atomic-resolution images of all four samples at 100 K and extracted the ellipticity (see Methods) of the Ge$|$Fe and Fe columns. In samples with long-range CDW (i.e., $A$ and $BA$), the Ge$|$Fe columns were more elliptical than the Fe columns, whereas no significant differences were observed between the two columns in the no-CDW samples ($B$ and $AB$) (Fig. 3q). Thus, our HAADF-STEM measurements suggest Ge displacement in the Fe$_3$Ge Kagome layer.

{\bf Muon Spin Rotation ($\mu$SR) measurements.}
To gain further insight into the magnetic properties of FeGe, we employed the $\mu$SR technique, which serves as an extremely sensitive local probe for detecting microscopic details of the static AFM order, ordered magnetic volume fraction, and magnetic fluctuations (Fig. 5a).  Figures 5d-f show the zero-field $\mu$SR time-spectra at different temperatures
for as-grown, no-CDW, and long-range CDW samples, respectively. At 410 K ($T>T_N$), the entire samples are in the paramagnetic state as evidenced by the weak $\mu$SR depolarization and its Gaussian functional form arising from the interaction between the muon spin and randomly oriented nuclear magnetic moments \cite{suter_musrfit_2012}. At 5 K, a spontaneous muon spin precession with a well-defined single frequency is observed (Figs. 4d-f). The behavior at intermediate temperatures in these three samples is quite different. 

Using the data analysis procedure described in the methods sections, we obtained quantitative information about the magnetic volume fraction and the magnitude of the static internal field at the muon site for each of these FeGe samples. Figure 5b shows the temperature dependence of the magnetically ordered fraction F in a wide temperature range. The magnetic volume fraction increases on cooling for all three samples. Around the CDW ordering temperature, no-CDW FeGe has the highest magnetic volume fraction, nearly 100 \%, consistent with STEM data in Fig. 3f,i, which shows no extended structural defects in the sample.
For comparison, magnetic volume fractions for long-range CDW and as-grown samples are lower, which we attribute to the formation of extended defects as shown in Figs. 3c,l. The temperature dependence of the internal fields ($\mu_0 H_{int}=\omega ⁄(\gamma_\mu^{-1} )$) for these three samples is shown in Figure 5c. In the as-grown and no-CDW samples, an internal field emerges just below 400 K, with a further increase observed around  150 K and the incommensurate AFM ordering temperature (30-50 K). This suggests that the total magnetic moment of the system increases below $T_{canting}$. Conversely, in the sample with long-range CDW, the internal field remains zero from below 400 K down to 100 K, after which it becomes detectable. This indicates that, in the long-range CDW sample, the magnetic order is relatively disordered between 400 K and 100 K on the $\mu$SR measurement time scale, and well-defined oscillations only appear below $T_{CDW}$.

\section{Discussion}\label{sec12}

Our comprehensive measurements provide a tantalizing microscopic scenario for the annealing process and demonstrate its impact on the macroscopic properties of kagome lattice antiferromagnet FeGe.
Using neutron polarization analysis, we have conclusively shown that CDW order below $\sim$100 K is coupled with an increased magnetic moment of iron in FeGe \cite{teng_discovery_2022}. Furthermore, we find that although the double-cone canted AFM
structure can describe the incommensurate magnetic peaks in as-grown samples \cite{bernhard_neutron_1984,bernhard_magnetic_1988}, it cannot account for the dramatic enhancement of the incommensurate magnetic scattering in the long-range CDW sample, suggesting the formation of a screw magnetic structure or spin density wave along the $c$-axis below $T_{canting}$ weakly coupled with the A-type AFM order (Fig. 1d) \cite{chen_competing_2024}. Instead of being associated with the CDW ordering temperature \cite{teng_discovery_2022}, the AHE in the spin-flop phase of the long-range CDW sample is dramatically enhanced compared with the as-grown sample 
and appears below $T_{canting}$ (Fig. 1g). Therefore, AHE in FeGe is likely to stem from itinerant electron interactions with incommensurate AFM order and field-induced noncollinear spin structure in the spin-flop phase \cite{wang_field-induced_2021,ghimire_competing_2020,dally_chiral_2021}.

From previous experimental and theoretical work, it was argued that the tunability of the CDW order in FeGe is closely intertwined with the disorder on the Ge1-site
\cite{wu_annealing-tunable_2024,chen_discovery_2024,shi_disordered_2023,wang_enhanced_2023,zhou_magnetic_2023,yi_polarized_2024,zhao_photoemission_2023}. From systematic mass measurements in different annealing processes (Table 1), we know that FeGe single crystals lose mass upon annealing at 560\textdegree C,
but not so when annealed at 320\textdegree C. Attributing all mass loss to Ge results in a loss of 1 in every 400 Ge atoms in the system for each annealing cycle, which is beyond the detection limits of EDX (Supplementary Fig. 7). The resulting Ge vacancies are distributed uniformly throughout the system for samples annealed at 560\textdegree C as indicated by the smooth BF-TEM images in Figs. 3f,i. Conversely, Ge vacancies coalesce in samples annealed at 320\textdegree C, forming dislocations and stacking faults, leaving the rest of FeGe sample to have lower point defect concentration (Fig. 3c,l). We note that the CDW order near defective regions is enhanced (Fig. 4a,b), thus suggesting that defects stabilize the long-range CDW order.  From the intensity line profiles of the Fe$_3$Ge layer for long-range and no-CDW samples (Fig. 3o), we conclude that Ge is lost primarily from the Ge1-sites. The uniform shortage of Ge1 atoms in no-CDW samples can be interpreted as a reduced occupancy at each Ge1 site. Therefore, for samples annealed at 560\textdegree C, the uniform distribution of Ge vacancies at Ge1-sites interrupts large-scale Ge-Ge dimerization, thereby preventing the formation of long-range CDW order. The coalescing of Ge vacancies to form stacking faults in samples annealed at 320\textdegree C yields regions in the sample with full occupancy at the Ge1 site (Fig. 1i) enabling the formation of Ge-Ge dimers and long-range CDW order consequently (Figs. 4a-d). 

The mechanism described above is similar to the microscopic annealing process of electron-doped copper oxides \cite{kang_microscopic_2007}. There, Cu vacancies in as-grown samples can be ``repaired" by removing oxygen in CuO$_2$ layers through the annealing process, resulting in one distinct region with fully occupied CuO$_2$ layers and another region with a C-type sesquioxide structured impurity phase without Cu. By reoxygenating superconducting electron-doped sample, both the superconductivity and impurity phase disappear \cite{kang_microscopic_2007}.

The enhancement of the AHE in long-range CDW samples in Fig. 1g can also be understood through a similar microscopic picture. Formation of the defect regions provides more efficient pathways for electrons, resulting in the lower residual resistivity ratio (RRR) for samples with long-range CDW order \cite{shi_disordered_2023}. This produces an exceptionally large anomalous Hall conductivity that is of similar magnitude to the giant AHE observed in KV$_3$Sb$_5$ \cite{yang_giant_2020,yu_concurrence_2021}. Since the onset of AHE in long-range CDW samples appears at temperatures far below $T_{CDW}$ (Fig. 1g), 
it cannot arise from the chiral flux phase as originally speculated \cite{teng_discovery_2022,feng_chiral_2021}.

In a recent work, it was found that long-range CDW order along the $c$-axis is suppressed below the incommensurate AFM ordering temperature while the in-plane CDW order persists \cite{shi_disordered_2023}. Although there is no evidence of intensity reduction at $(0,0,0.5)$ across $T_{canting}$ in our long-range CDW sample (Fig. 2k), we note that neutron scattering is not sensitive to electron cloud and only sensitive to nuclei-induced lattice distortions. It is possible that a small charge-redistribution below $T_{canting}$ does not dramatically affect the nuclei positions.
 An X-ray diffuse scattering experiment suggests that
the CDW order in FeGe is the order-disorder type originating from  
the frustration introduced by the dimerization of the original G1-site on the kagome plane \cite{subires_frustrated_2024}, consistent with our notion that the dimerization in Ge1 vacancy-free regions is critical for the CDW transition. In kagome lattice FeGe, the annealing tunable CDW order and the resulting dramatic changes in electronic, magnetic, and transport properties make it a unique quantum material potentially useful for in-situ annealed controlled sensing and information transmission.

\section{Methods}\label{sec11}
\subsection{FeGe Synthesis and Annealing}
Single crystals of FeGe were synthesized via vapor chemical transport as described in previous work \cite{teng_discovery_2022}. In addition to the as-grown samples, three other categories of single crystals were subject to different post-growth annealing procedures which include samples with long-range CDW order, short-range CDW order, and no-CDW order. Long-range CDW samples were annealed in a vacuum at 320\textdegree C for 96 hours and then quenched immediately with water, as described in \cite{wu_annealing-tunable_2024}. Short-range CDW samples were prepared under the same conditions, but with an annealing temperature of 320\textdegree C for 8 hours. Lastly, samples with no-CDW order were annealed at 560\textdegree C for 96 hours.  

A collection of single crystals was cycled six times between 560\textdegree C for 96 hours and 320\textdegree C for 96 hours to show the loss and recovery of long-range CDW order. The mass of the samples was recorded on a Sartorius MSE-125P-100-DA electronic scale between each annealing with a tolerance of 0.01 mg.      
\subsection{Neutron Scattering}
Unpolarized neutron scattering measurements on FeGe were carried out on the CORELLI, BL-9, spectrometer \cite{ye_implementation_2018} of the Spallation Neutron Source and the HB-3A DEMAND \cite{cryst9010005} and HB-1A spectrometers at the High Flux Isotope Reactor (HFIR) at Oak Ridge National Lab (ORNL), USA. Neutron polarization analysis \cite{liu_-plane_2020,lipscombe_anisotropic_2010} was performed on the HB-1 polarized triple-axis spectrometer at HFIR. To directly compare the impact of post-growth annealing, all neutron scattering data on the long-range CDW sample was collected using the same 30 mg single crystal used in \cite{teng_discovery_2022}. Specifically, all neutron data in Fig 2 labeled “long-range CDW” and “as-grown” was first collected on the as-grown sample, then annealed at 320\textdegree C for 96 hours and measured again. The short-range CDW and no-CDW samples are distinct and roughly 15 mg each, but from the same batch as the 30 mg sample. The momentum transfer \boldmath$Q$ in 3D reciprocal space in \r{A}\textsuperscript{-1} was defined as \unboldmath$\pmb{Q} = H\pmb{a^*}+K\pmb{b^*}+L\pmb{c^*}$ where $H$, $K$, and $L$ are Miller indices and $\pmb{a^*}=2\pi(\pmb{b} \times \pmb{c})/[\pmb{a} \cdot (\pmb{b} \times \pmb{c})]$, $\pmb{b^*}=2\pi(\pmb{c} \times \pmb{a})/[\pmb{b} \cdot (\pmb{c} \times \pmb{a})]$, $\pmb{c^*}=2\pi(\pmb{a} \times \pmb{b})/[\pmb{c} \cdot (\pmb{a} \times \pmb{b})]$ with $\pmb{a}=a\pmb{\hat{x}}$, $\pmb{b}=a(\cos 120\pmb{\hat{x}}+\sin 120\pmb{\hat{y}})$, and $\pmb{c}=c\pmb{\hat{z}}$ ($a \approx b \approx 4.99$\r{A}, $c \approx 4.05$\r{A} at room temperature).
\subsection{Electrical transport and magnetic susceptibility}
Electrical resistivity and Hall effect were simultaneously studied using a Quantum Design physical property measurement system (PPMS) with five-point contacts. A magnetic field up to 14 T was applied parallel to the $c$-axis, while the current was applied in the $ab$ plane. Magnetic susceptibility measurements were conducted using the VSM option on the PPMS.

\subsection{Structural and Magnetic Refinement}
To determine the structural evolution of FeGe across the CDW transition, we performed crystal structure refinement using the Fullprof Suite software. 3D reciprocal space maps of FeGe were collected at CORELLI above and below $T_{CDW}$ for all three annealing conditions. Integrated intensities of the structural Bragg peaks were extracted for use in the refinement process. Additionally, a separate set of integrated intensities was collected on the four-circle diffractometer HB-3A for the long-range and short-range CDW samples. We utilized the dimerization framework described theoretically and experimentally in earlier work to construct a $2\times2\times2$ unit cell to refine the positions of corner Ge1 atoms. 

The canted AFM phase contains two propagation vectors, (0,0,0.46) and (0,0,0.5). The magnetic intensities of the double cone AFM structure at low temperatures can be represented by the following equation \cite{bernhard_magnetic_1988}: 

\begin{multline}
F^2(\pmb{q})=A(f_{Fe}(\pmb{q}))^2 |G_{HKL}|^2 [(1-e_z^2)\delta (\pmb{q}-\tau _{HKL})\cos^2 \alpha 
\\+
\frac{1}{4}(1+e_z^2)\delta (\pmb{q}+\pmb{Q}-\tau _{HKL})+\delta (\pmb{q}-\pmb{Q}-\tau _{HKL})\sin^2 \alpha]
\label{eq1}
\end{multline}

where $A$ is a constant scaling factor, $f_{Fe}$ is the magnetic form factor of iron, $G_{HKL}$ is the geometric structure factor for the Fe kagome lattice where $G_{HKL} = 6$ for even $H$ and $G_{HKL}=2$ for odd $H$ when restricted to the $[H,0,L]$ scattering plane, $e_z$ is the z-axis projection of the unit scattering vector $q$, $\tau _{HKL}$ is the reciprocal lattice vector for the $(H,K,L)$ reflection, $\pmb{Q}=(0,0,0.04)$ is the modulation vector, and $\alpha$ is the canting angle. The only unknown variable is the canting angle $\alpha$ which we determine through refinement. Each annealing condition possesses stark differences in the relative intensities of incommensurate Bragg peaks (Fig. 2o). Setting the canting angle to 16\textdegree \  and 28\textdegree \ for the as-grown and long-range CDW samples respectively results in strong agreement with the experimental data (Fig. 2(l-m)). 

\subsection{Neutron Polarization Analysis}
The triple-axis spectrometer HB-1 at ORNL was used to conduct polarized neutron scattering experiments. A vertically focused Heusler monochrometer was used to select neutrons with particular polarizations. To detect the polarized neutrons, a flat Heusler analyzer was used. As neutrons are not sensitive to magnetic scattering unless they are perpendicular to $\pmb{Q}$ \cite{boothroyd_principles_2020}, defining the $x$ direction as parallel to $\pmb{Q}$ allows for magnetic responses in the $y-z$ plane to be measured. Scattered neutrons may have parallel or antiparallel spins (NSF and SF respectively) with respect to the incident neutrons. A neutron will see its spin flipped only if the incident spin is perpendicular to the moment within the sample. Solving for $\sigma_{x}^{SF}$, $\sigma_{y}^{SF}$, and $\sigma_{z}^{SF}$ using the equation in the main text, we find 

\begin{equation}
\begin{split}
\sigma_{x}^{SF}=\frac{R}{R+1}M_y+\frac{R}{R+1}M_z+B,\\
\sigma_{y}^{SF}=\frac{1}{R+1}M_y+\frac{R}{R+1}M_z+B,\\
\sigma_{z}^{SF}=\frac{R}{R+1}M_y+\frac{1}{R+1}M_z+B
\end{split}
\end{equation}

where $B$ is a combination of the sample environment background $b_1$, nuclear scattering $N$, and $\pmb{Q}$-independent incoherent scattering. We can simplify this further by taking the difference of scattering cross sections 

\begin{equation}
\begin{split}
\sigma_{x}^{SF} - \sigma_{y}^{SF} = \frac{R-1}{R+1}M_y,\\
\sigma_{x}^{SF} - \sigma_{z}^{SF} = \frac{R-1}{R+1}M_z
\end{split}
\end{equation}

To determine the direction of the AFM order, recall $M_y=\sin ^2(\theta) M_{b\perp}+\cos ^2(\theta)M_c$. At the AFM Bragg peak (2,0,0.5), $\theta \approx 17$\textdegree. The equation for $M_y$ becomes $M_y\propto 0.09M_{b\perp}+0.91M_c$, meaning $M_y$ is incredibly sensitive to moments along the c-axis. By contrast, $M_z \propto \sigma_{x}^{SF} - \sigma_{z}^{SF}$ and is highly sensitive to moments perpendicular to the c-axis. As $\sigma_{x}^{SF} \approx \sigma_{z}^{SF}$ from Fig. 2g, we conclude that there is virtually no magnetic contribution to $M_z$. Additionally, $\sigma_y^{SF} \approx \sigma_x^{NSF}$ meaning there is no magnetic signal picked up by $\sigma_y^{SF}$, in effect maximizing $M_y$. From this we conclude that the moment direction of the AFM order is exclusively along the c-axis.

\subsection{Scanning Transmission Electron Microscopy}
Cross-section samples, along $[110]$ zone axis, were prepared from single crystals of FeGe using a Ga focused-ion beam (FIB) liftout procedure on a Thermo Fisher Scientific (TFS) Helios G4 X FIB. (S)TEM data at 100 K were collected on a FEI/TFS Titan Themis 300 and a TFS Spectra 300 X-CFEG at an accelerating voltage of 300 kV, with a Gatan 636 double-tilt cryogenic sample holder. 
Atomic-resolution STEM images were acquired using a high-angle annular dark-field (HAADF) detector with a convergence semi-angle of 30 mrad, and beam current of 50-70 pA. To avoid mechanical and thermal instabilities in atomic-resolution images at cryogenic temperatures, multiple images were acquired with a frame rate of 0.26 s, and subsequently registered to get a high signal-to-noise ratio image \cite{savitzky_image_2018}. The mean unit cells presented in Fig. 3e,h,k,n are obtained from the registered images by measuring the shape and position of the atomic columns after fitting two-dimensional (2D) Gaussians \cite{funni_theory_2021}. Intensities for each identified atomic column were averaged to obtain a mean unit cell.

The form of the 2D Gaussian used for fitting is given by:
\begin{equation}
    G(x,y)=A e^{\left[-\frac{1}{2} \left\{ \left(\frac{(x-x_o)\cos \theta + (y-y_o)\sin \theta}{\sigma_1} \right) ^2 + \left(\frac{-(x-x_o)\sin \theta + (y-y_o)\cos \theta}{\sigma_2}\right)^2 \right\} \right]}
\end{equation}
where $A$ is the amplitude of the Gaussian, ($x_o, y_o$) is the center, $\sigma_1, \sigma_2$ are the spread (standard deviations) of the Gaussian along two principal axes, and $\theta$ is the rotation angle determining the orientation of the Gaussian.

After fitting each atomic column with a 2D Gaussian, the ellipticities were extracted by $e = \frac{\sigma_{major}}{\sigma_{minor}}$, where $\sigma_{major} = \sigma_1, \sigma_{minor} = \sigma_2$ if $\sigma_1 > \sigma_2$, else $\sigma_{major} = \sigma_2, \sigma_{minor} = \sigma_1$.

 The simulated HAADF-STEM image of FeGe in [110] zone axis was generated using abTEM’s multislice package \cite{madsen_abtem_2021} using 300 keV electrons and a 30 mrad convergence semi-angle. 25 frozen phonons were used to simulate the effect of thermal vibrations. 
The 4D-STEM datasets were acquired using an EMPAD-G2 detector with a beam current of $\sim$ 40 pA and convergence semi-angle of 0.2 mrad (probe size of $\sim$ 7 nm). To obtain the CDW maps, each diffraction pattern in the dataset was upsampled by a factor of 4. The Bragg peaks and CDW peaks were identified, and their positions refined by fitting 2D Gaussians. For each of the real-space map, two sets of virtual apertures were defined as binary masks: one for the Bragg or CDW peaks, and the other for region around the diffraction peaks for local background intensity.  The real-space Bragg and CDW maps were constructed by integrating the intensity within the virtual aperture for the diffraction peaks. The mean background is subtracted before the integration.

\subsection{Muon Spin Rotation}
In a $\mu$SR experiment nearly 100\% spin-polarized muons $\mu^+$ are implanted into the sample one at a time. The positively charged $\mu^+$ thermalize at interstitial lattice sites, where they act as magnetic microprobes. In a magnetic material, the muon spin precesses in the local field $B$ at the muon site with the Larmor frequency (muon gyromagnetic ratio $\frac{\gamma _\mu}{2\pi}=$135.5 MHz T$^{-1}$). 

The GPS ($\pi$M3 beamline) $\mu$SR instrument \cite{amato_new_2017}, equipped with He gas-flow cryostat and CCR at the Paul Scherrer Institute, Switzerland, was used to study the single crystalline samples of FeGe. Single crystals used in $\mu$SR measurements are from a separate batch of samples than the single crystal used in our neutron scattering experiments. Measurements are conducted under zero-field conditions to obtain information about the internal magnetic field and the fraction of ordered magnetic volume. The ZF-$\mu$SR spectra for x $<$ 0.3 were analyzed using the following incommensurate model:
\begin{multline}
    A_{ZF}(t)=F\left( \sum\limits_{j=1}^2 \left( f_j J_0 (\gamma_{\mu} B_{int} t)e^{-\lambda_{j}t} \right) + f_Le^{-\lambda_{L}t}\right) \\+ (1-F) \left(\frac{1}{3} + \frac{2}{3}\left( 1-(\sigma t)^2 \right) e^{-\frac{1}{2}(\sigma t)^2}\right)
\end{multline}

This model consists of an anisotropic magnetic contribution characterized by an oscillating “transverse” component and a slowly relaxing “longitudinal” component. The longitudinal component arises due to muons experiencing local field components which are parallel to the initial muon spin polarization. In polycrystalline samples with randomly oriented grains the orientational averaging results in a so-called “one-third tail” with $f_L$ = 1/3. For single crystals, $f_L$ varies between zero and unity as the orientation between field and polarization changes from being perpendicular to parallel. $J_0$ is the zeroth order Bessel function which is used to describe the field distribution created by incommensurate magnetism. The depolarization rates $\lambda_T$ and $\lambda_L$ characterize the damping of the oscillating and non-oscillating part of the $\mu$SR signal, respectively. Note that in the entire investigated temperature range, in addition to the magnetically ordered contribution, there is a paramagnetic signal component characterized by the width $\sigma$ of the field distribution at the muon site created by the densely distributed nuclear moments. The temperature-dependent magnetic ordering fraction $0 \leq F \leq 1$ governs the trade-off between magnetically-ordered and paramagnetic behaviors.

\backmatter

\bmhead{Acknowledgements}
C.-L.H. acknowledges Dr. M.K. Lee at the PPMS-16T Lab, Instrumentation Center, National Cheng Kung University (NCKU) for technical support. Y.-C.C. and C.-L.H. are grateful to P.-Z. Hsu and L.-J. Chang at NCKU for the help of Laue diffraction. 
The neutron scattering and single crystal synthesis work at Rice are supported by US NSF DMR-2401084 and the Robert A. Welch Foundation under Grant No. C-1839, respectively (P.D.). The STEM and 4D STEM characterization at cryogenic temperature performed at Cornell University is supported by US DOE BES DE-SC0023905 (J.J.C). The electron microscopy facilities are supported by the NSF (Platform for the Accelerated Realization, Analysis, and Discovery of Interface Materials) under Cooperative Agreement No. DMR-2039380, and the Cornell Center for Materials Research shared instrument facilities. The Titan Themis 300 was acquired through NSF-MRI-1429155, with additional support from Cornell University, the Weill Institute and the Kavli Institute at Cornell. Z.G. acknowledges support from the Swiss National Science Foundation (SNSF) through SNSF Starting Grant (number TMSGI2\_211750).  Y.-C.C. and C.-L.H. acknowledge support by the National Science and Technology Council in Taiwan, NSTC 113-2112-M-006-027, and the Higher Education Sprout Project, Ministry of Education to the Headquarters of University Advancement at NCKU.  D.N. and T.L.J. are
supported by US NSF DMR-2102028. H.C. and Y.Q.H. acknowledge support from the U.S. DOE, BES, Early Career Research Program Award KC0402020,
under Contract DE-AC05-00OR22725. 
M.Y. was supported by the U.S. DOE grant No. DE-SC0021421, the Gordon and Betty Moore Foundation's EPiQS Initiative through grant No. GBMF9470 and the Robert A. Welch Foundation Grant No. C-2175.
This research used resources at the High Flux Isotope Reactor and Spallation Neutron Source,  DOE Office of Science User Facilities operated by the Oak Ridge National Laboratory.

\section*{Declarations}

\subsection{Author contributions} P.D., M.Y., and J.J.C. designed and managed the research.
The sample preparation, annealing, and transport measurements were carried out by M.L.K, Y.C.C., S.j.X., Y.F.X., T.L., D.N., and C.L.H.. Neutron scattering experiments and analysis were carried out by M.L.K., F.Y., H.C., Y.Q.H., W.T., M.M., and P.D.. STEM measurements and analysis were performed by S.S., M.T.K., and J.J.C.. $\mu$SR measurements and analysis were done by H.L. and Z.G.. The paper was written by M.L.K, S.S., Z.G., J.J.C., and P.D. with comments from all co-authors. 

\subsection{Competing interests} The authors declare no competing interests.

\noindent


\bibliography{sn-bibliography}


\begin{thebibliography}{59}
\ifx \bisbn   \undefined \def \bisbn  #1{ISBN #1}\fi
\ifx \binits  \undefined \def \binits#1{#1}\fi
\ifx \bauthor  \undefined \def \bauthor#1{#1}\fi
\ifx \batitle  \undefined \def \batitle#1{#1}\fi
\ifx \bjtitle  \undefined \def \bjtitle#1{#1}\fi
\ifx \bvolume  \undefined \def \bvolume#1{\textbf{#1}}\fi
\ifx \byear  \undefined \def \byear#1{#1}\fi
\ifx \bissue  \undefined \def \bissue#1{#1}\fi
\ifx \bfpage  \undefined \def \bfpage#1{#1}\fi
\ifx \blpage  \undefined \def \blpage #1{#1}\fi
\ifx \burl  \undefined \def \burl#1{\textsf{#1}}\fi
\ifx \doiurl  \undefined \def \doiurl#1{\url{https://doi.org/#1}}\fi
\ifx \betal  \undefined \def \betal{\textit{et al.}}\fi
\ifx \binstitute  \undefined \def \binstitute#1{#1}\fi
\ifx \binstitutionaled  \undefined \def \binstitutionaled#1{#1}\fi
\ifx \bctitle  \undefined \def \bctitle#1{#1}\fi
\ifx \beditor  \undefined \def \beditor#1{#1}\fi
\ifx \bpublisher  \undefined \def \bpublisher#1{#1}\fi
\ifx \bbtitle  \undefined \def \bbtitle#1{#1}\fi
\ifx \bedition  \undefined \def \bedition#1{#1}\fi
\ifx \bseriesno  \undefined \def \bseriesno#1{#1}\fi
\ifx \blocation  \undefined \def \blocation#1{#1}\fi
\ifx \bsertitle  \undefined \def \bsertitle#1{#1}\fi
\ifx \bsnm \undefined \def \bsnm#1{#1}\fi
\ifx \bsuffix \undefined \def \bsuffix#1{#1}\fi
\ifx \bparticle \undefined \def \bparticle#1{#1}\fi
\ifx \barticle \undefined \def \barticle#1{#1}\fi
\bibcommenthead
\ifx \bconfdate \undefined \def \bconfdate #1{#1}\fi
\ifx \botherref \undefined \def \botherref #1{#1}\fi
\ifx \url \undefined \def \url#1{\textsf{#1}}\fi
\ifx \bchapter \undefined \def \bchapter#1{#1}\fi
\ifx \bbook \undefined \def \bbook#1{#1}\fi
\ifx \bcomment \undefined \def \bcomment#1{#1}\fi
\ifx \oauthor \undefined \def \oauthor#1{#1}\fi
\ifx \citeauthoryear \undefined \def \citeauthoryear#1{#1}\fi
\ifx \endbibitem  \undefined \def \endbibitem {}\fi
\ifx \bconflocation  \undefined \def \bconflocation#1{#1}\fi
\ifx \arxivurl  \undefined \def \arxivurl#1{\textsf{#1}}\fi
\csname PreBibitemsHook\endcsname

\bibitem[\protect\citeauthoryear{Keimer et~al.}{2015}]{keimer_quantum_2015}
\begin{barticle}
\bauthor{\bsnm{Keimer}, \binits{B.}},
\bauthor{\bsnm{Kivelson}, \binits{S.A.}},
\bauthor{\bsnm{Norman}, \binits{M.R.}},
\bauthor{\bsnm{Uchida}, \binits{S.}},
\bauthor{\bsnm{Zaanen}, \binits{J.}}:
\batitle{From quantum matter to high-temperature superconductivity in copper oxides}.
\bjtitle{NATURE}
\bvolume{518}(\bissue{7538}),
\bfpage{179}--\blpage{186}
(\byear{2015})
\doiurl{10.1038/nature14165}
\end{barticle}
\endbibitem

\bibitem[\protect\citeauthoryear{Armitage et~al.}{2010}]{armitage_progress_2010}
\begin{barticle}
\bauthor{\bsnm{Armitage}, \binits{N.P.}},
\bauthor{\bsnm{Fournier}, \binits{P.}},
\bauthor{\bsnm{Greene}, \binits{R.L.}}:
\batitle{Progress and perspectives on electron-doped cuprates}.
\bjtitle{Rev. Mod. Phys.}
\bvolume{82}(\bissue{3}),
\bfpage{2421}--\blpage{2487}
(\byear{2010})
\doiurl{10.1103/RevModPhys.82.2421} .
\bcomment{Publisher: American Physical Society}
\end{barticle}
\endbibitem

\bibitem[\protect\citeauthoryear{Dai}{2015}]{dai_antiferromagnetic_2015}
\begin{barticle}
\bauthor{\bsnm{Dai}, \binits{P.}}:
\batitle{Antiferromagnetic order and spin dynamics in iron-based superconductors}.
\bjtitle{Rev. Mod. Phys.}
\bvolume{87}(\bissue{3}),
\bfpage{855}--\blpage{896}
(\byear{2015})
\doiurl{10.1103/RevModPhys.87.855} .
\bcomment{Publisher: American Physical Society}
\end{barticle}
\endbibitem

\bibitem[\protect\citeauthoryear{Fradkin et~al.}{2015}]{fradkin_colloquium_2015}
\begin{barticle}
\bauthor{\bsnm{Fradkin}, \binits{E.}},
\bauthor{\bsnm{Kivelson}, \binits{S.A.}},
\bauthor{\bsnm{Tranquada}, \binits{J.M.}}:
\batitle{Colloquium: {Theory} of intertwined orders in high temperature superconductors}.
\bjtitle{Rev. Mod. Phys.}
\bvolume{87}(\bissue{2}),
\bfpage{457}--\blpage{482}
(\byear{2015})
\doiurl{10.1103/RevModPhys.87.457} .
\bcomment{Publisher: American Physical Society}
\end{barticle}
\endbibitem

\bibitem[\protect\citeauthoryear{TOKURA et~al.}{1989}]{tokura_superconducting_1989}
\begin{barticle}
\bauthor{\bsnm{TOKURA}, \binits{Y.}},
\bauthor{\bsnm{TAKAGI}, \binits{H.}},
\bauthor{\bsnm{UCHIDA}, \binits{S.}}:
\batitle{A {SUPERCONDUCTING} {COPPER}-{OXIDE} {COMPOUND} {WITH} {ELECTRONS} {AS} {THE} {CHARGE}-{CARRIERS}}.
\bjtitle{NATURE}
\bvolume{337}(\bissue{6205}),
\bfpage{345}--\blpage{347}
(\byear{1989})
\doiurl{10.1038/337345a0}
\end{barticle}
\endbibitem

\bibitem[\protect\citeauthoryear{Takagi et~al.}{1989}]{takagi_superconductivity_1989}
\begin{barticle}
\bauthor{\bsnm{Takagi}, \binits{H.}},
\bauthor{\bsnm{Uchida}, \binits{S.}},
\bauthor{\bsnm{Tokura}, \binits{Y.}}:
\batitle{Superconductivity produced by electron doping in {\textbackslash}{mathrmCuO}\_2-layered compounds}.
\bjtitle{Phys. Rev. Lett.}
\bvolume{62}(\bissue{10}),
\bfpage{1197}--\blpage{1200}
(\byear{1989})
\doiurl{10.1103/PhysRevLett.62.1197} .
\bcomment{Publisher: American Physical Society}
\end{barticle}
\endbibitem

\bibitem[\protect\citeauthoryear{Higgins et~al.}{2006}]{higgins_role_2006}
\begin{barticle}
\bauthor{\bsnm{Higgins}, \binits{J.S.}},
\bauthor{\bsnm{Dagan}, \binits{Y.}},
\bauthor{\bsnm{Barr}, \binits{M.C.}},
\bauthor{\bsnm{Weaver}, \binits{B.D.}},
\bauthor{\bsnm{Greene}, \binits{R.L.}}:
\batitle{Role of oxygen in the electron-doped superconducting cuprates}.
\bjtitle{Phys. Rev. B}
\bvolume{73}(\bissue{10}),
\bfpage{104510}
(\byear{2006})
\doiurl{10.1103/PhysRevB.73.104510} .
\bcomment{Publisher: American Physical Society}
\end{barticle}
\endbibitem

\bibitem[\protect\citeauthoryear{Kang et~al.}{2007}]{kang_microscopic_2007}
\begin{barticle}
\bauthor{\bsnm{Kang}, \binits{H.J.}},
\bauthor{\bsnm{Dai}, \binits{P.}},
\bauthor{\bsnm{Campbell}, \binits{B.J.}},
\bauthor{\bsnm{Chupas}, \binits{P.J.}},
\bauthor{\bsnm{Rosenkranz}, \binits{S.}},
\bauthor{\bsnm{Lee}, \binits{P.L.}},
\bauthor{\bsnm{Huang}, \binits{Q.}},
\bauthor{\bsnm{Li}, \binits{S.}},
\bauthor{\bsnm{Komiya}, \binits{S.}},
\bauthor{\bsnm{Ando}, \binits{Y.}}:
\batitle{Microscopic annealing process and its impact on superconductivity in {T}-structure electron-doped copper oxides}.
\bjtitle{Nature Materials}
\bvolume{6}(\bissue{3}),
\bfpage{224}--\blpage{229}
(\byear{2007})
\doiurl{10.1038/nmat1847} .
Accessed 2024-06-26
\end{barticle}
\endbibitem

\bibitem[\protect\citeauthoryear{Guarino et~al.}{2022}]{guarino_superconductivity_2022}
\begin{barticle}
\bauthor{\bsnm{Guarino}, \binits{A.}},
\bauthor{\bsnm{Autieri}, \binits{C.}},
\bauthor{\bsnm{Marra}, \binits{P.}},
\bauthor{\bsnm{Leo}, \binits{A.}},
\bauthor{\bsnm{Grimaldi}, \binits{G.}},
\bauthor{\bsnm{Avella}, \binits{A.}},
\bauthor{\bsnm{Nigro}, \binits{A.}}:
\batitle{Superconductivity induced by structural reorganization in the electron-doped cuprate {\textbackslash}{mathrmNd}\_2{\textbackslash}ensuremath-x{\textbackslash}{mathrmCe}\_x{\textbackslash}{mathrmCuO}\_4}.
\bjtitle{Phys. Rev. B}
\bvolume{105}(\bissue{1}),
\bfpage{014512}
(\byear{2022})
\doiurl{10.1103/PhysRevB.105.014512} .
\bcomment{Publisher: American Physical Society}
\end{barticle}
\endbibitem

\bibitem[\protect\citeauthoryear{Wu et~al.}{2024}]{wu_reversible_2024}
\begin{botherref}
\oauthor{\bsnm{Wu}, \binits{H.}},
\oauthor{\bsnm{Chen}, \binits{L.}},
\oauthor{\bsnm{Malinowski}, \binits{P.}},
\oauthor{\bsnm{Jang}, \binits{B.G.}},
\oauthor{\bsnm{Deng}, \binits{Q.}},
\oauthor{\bsnm{Scott}, \binits{K.}},
\oauthor{\bsnm{Huang}, \binits{J.}},
\oauthor{\bsnm{Ruff}, \binits{J.P.C.}},
\oauthor{\bsnm{He}, \binits{Y.}},
\oauthor{\bsnm{Chen}, \binits{X.}},
\oauthor{\bsnm{Hu}, \binits{C.}},
\oauthor{\bsnm{Yue}, \binits{Z.}},
\oauthor{\bsnm{Oh}, \binits{J.S.}},
\oauthor{\bsnm{Teng}, \binits{X.}},
\oauthor{\bsnm{Guo}, \binits{Y.}},
\oauthor{\bsnm{Klemm}, \binits{M.}},
\oauthor{\bsnm{Shi}, \binits{C.}},
\oauthor{\bsnm{Shi}, \binits{Y.}},
\oauthor{\bsnm{Setty}, \binits{C.}},
\oauthor{\bsnm{Werner}, \binits{T.}},
\oauthor{\bsnm{Hashimoto}, \binits{M.}},
\oauthor{\bsnm{Lu}, \binits{D.}},
\oauthor{\bsnm{Yilmaz}, \binits{T.}},
\oauthor{\bsnm{Vescovo}, \binits{E.}},
\oauthor{\bsnm{Mo}, \binits{S.-K.}},
\oauthor{\bsnm{Fedorov}, \binits{A.}},
\oauthor{\bsnm{Denlinger}, \binits{J.D.}},
\oauthor{\bsnm{Xie}, \binits{Y.}},
\oauthor{\bsnm{Gao}, \binits{B.}},
\oauthor{\bsnm{Kono}, \binits{J.}},
\oauthor{\bsnm{Dai}, \binits{P.}},
\oauthor{\bsnm{Han}, \binits{Y.}},
\oauthor{\bsnm{Xu}, \binits{X.}},
\oauthor{\bsnm{Birgeneau}, \binits{R.J.}},
\oauthor{\bsnm{Zhu}, \binits{J.-X.}},
\oauthor{\bsnm{Silva~Neto}, \binits{E.H.}},
\oauthor{\bsnm{Wu}, \binits{L.}},
\oauthor{\bsnm{Chu}, \binits{J.-H.}},
\oauthor{\bsnm{Si}, \binits{Q.}},
\oauthor{\bsnm{Yi}, \binits{M.}}:
Reversible non-volatile electronic switching in a near-room-temperature van der {Waals} ferromagnet.
NATURE COMMUNICATIONS
\textbf{15}(1)
(2024)
\doiurl{10.1038/s41467-024-46862-z}
\end{botherref}
\endbibitem

\bibitem[\protect\citeauthoryear{Ohoyama et~al.}{1963}]{ohoyama_new_1963}
\begin{barticle}
\bauthor{\bsnm{Ohoyama}, \binits{T.}},
\bauthor{\bsnm{Kanematsu}, \binits{K.}},
\bauthor{\bsnm{Yasukōchi}, \binits{K.}}:
\batitle{A {New} {Intermetallic} {Compound} {FeGe}}.
\bjtitle{Journal of the Physical Society of Japan}
\bvolume{18}(\bissue{4}),
\bfpage{589}--\blpage{589}
(\byear{1963})
\doiurl{10.1143/JPSJ.18.589} .
Accessed 2024-07-02
\end{barticle}
\endbibitem

\bibitem[\protect\citeauthoryear{Bernhard et~al.}{1984}]{bernhard_neutron_1984}
\begin{barticle}
\bauthor{\bsnm{Bernhard}, \binits{J.}},
\bauthor{\bsnm{Lebech}, \binits{B.}},
\bauthor{\bsnm{Beckman}, \binits{O.}}:
\batitle{Neutron diffraction studies of the low-temperature magnetic structure of hexagonal {FeGe}}.
\bjtitle{Journal of Physics F: Metal Physics}
\bvolume{14}(\bissue{10}),
\bfpage{2379}
(\byear{1984})
\doiurl{10.1088/0305-4608/14/10/017}
\end{barticle}
\endbibitem

\bibitem[\protect\citeauthoryear{Bernhard et~al.}{1988}]{bernhard_magnetic_1988}
\begin{barticle}
\bauthor{\bsnm{Bernhard}, \binits{J.}},
\bauthor{\bsnm{Lebech}, \binits{B.}},
\bauthor{\bsnm{Beckman}, \binits{O.}}:
\batitle{Magnetic phase diagram of hexagonal {FeGe} determined by neutron diffraction}.
\bjtitle{Journal of Physics F: Metal Physics}
\bvolume{18}(\bissue{3}),
\bfpage{539}--\blpage{552}
(\byear{1988})
\doiurl{10.1088/0305-4608/18/3/023} .
Accessed 2024-06-26
\end{barticle}
\endbibitem

\bibitem[\protect\citeauthoryear{Teng et~al.}{2022}]{teng_discovery_2022}
\begin{barticle}
\bauthor{\bsnm{Teng}, \binits{X.}},
\bauthor{\bsnm{Chen}, \binits{L.}},
\bauthor{\bsnm{Ye}, \binits{F.}},
\bauthor{\bsnm{Rosenberg}, \binits{E.}},
\bauthor{\bsnm{Liu}, \binits{Z.}},
\bauthor{\bsnm{Yin}, \binits{J.-X.}},
\bauthor{\bsnm{Jiang}, \binits{Y.-X.}},
\bauthor{\bsnm{Oh}, \binits{J.S.}},
\bauthor{\bsnm{Hasan}, \binits{M.Z.}},
\bauthor{\bsnm{Neubauer}, \binits{K.J.}},
\bauthor{\bsnm{Gao}, \binits{B.}},
\bauthor{\bsnm{Xie}, \binits{Y.}},
\bauthor{\bsnm{Hashimoto}, \binits{M.}},
\bauthor{\bsnm{Lu}, \binits{D.}},
\bauthor{\bsnm{Jozwiak}, \binits{C.}},
\bauthor{\bsnm{Bostwick}, \binits{A.}},
\bauthor{\bsnm{Rotenberg}, \binits{E.}},
\bauthor{\bsnm{Birgeneau}, \binits{R.J.}},
\bauthor{\bsnm{Chu}, \binits{J.-H.}},
\bauthor{\bsnm{Yi}, \binits{M.}},
\bauthor{\bsnm{Dai}, \binits{P.}}:
\batitle{Discovery of charge density wave in a kagome lattice antiferromagnet}.
\bjtitle{Nature}
\bvolume{609}(\bissue{7927}),
\bfpage{490}--\blpage{495}
(\byear{2022})
\doiurl{10.1038/s41586-022-05034-z} .
Accessed 2024-06-26
\end{barticle}
\endbibitem

\bibitem[\protect\citeauthoryear{Yin et~al.}{2022}]{yin_discovery_2022}
\begin{barticle}
\bauthor{\bsnm{Yin}, \binits{J.-X.}},
\bauthor{\bsnm{Jiang}, \binits{Y.-X.}},
\bauthor{\bsnm{Teng}, \binits{X.}},
\bauthor{\bsnm{Hossain}, \binits{M.S.}},
\bauthor{\bsnm{Mardanya}, \binits{S.}},
\bauthor{\bsnm{Chang}, \binits{T.-R.}},
\bauthor{\bsnm{Ye}, \binits{Z.}},
\bauthor{\bsnm{Xu}, \binits{G.}},
\bauthor{\bsnm{Denner}, \binits{M.M.}},
\bauthor{\bsnm{Neupert}, \binits{T.}},
\bauthor{\bsnm{Lienhard}, \binits{B.}},
\bauthor{\bsnm{Deng}, \binits{H.-B.}},
\bauthor{\bsnm{Setty}, \binits{C.}},
\bauthor{\bsnm{Si}, \binits{Q.}},
\bauthor{\bsnm{Chang}, \binits{G.}},
\bauthor{\bsnm{Guguchia}, \binits{Z.}},
\bauthor{\bsnm{Gao}, \binits{B.}},
\bauthor{\bsnm{Shumiya}, \binits{N.}},
\bauthor{\bsnm{Zhang}, \binits{Q.}},
\bauthor{\bsnm{Cochran}, \binits{T.A.}},
\bauthor{\bsnm{Multer}, \binits{D.}},
\bauthor{\bsnm{Yi}, \binits{M.}},
\bauthor{\bsnm{Dai}, \binits{P.}},
\bauthor{\bsnm{Hasan}, \binits{M.Z.}}:
\batitle{Discovery of {Charge} {Order} and {Corresponding} {Edge} {State} in {Kagome} {Magnet} {FeGe}}.
\bjtitle{Phys. Rev. Lett.}
\bvolume{129}(\bissue{16}),
\bfpage{166401}
(\byear{2022})
\doiurl{10.1103/PhysRevLett.129.166401} .
\bcomment{Publisher: American Physical Society}
\end{barticle}
\endbibitem

\bibitem[\protect\citeauthoryear{Teng et~al.}{2023}]{teng_magnetism_2023}
\begin{barticle}
\bauthor{\bsnm{Teng}, \binits{X.}},
\bauthor{\bsnm{Oh}, \binits{J.S.}},
\bauthor{\bsnm{Tan}, \binits{H.}},
\bauthor{\bsnm{Chen}, \binits{L.}},
\bauthor{\bsnm{Huang}, \binits{J.}},
\bauthor{\bsnm{Gao}, \binits{B.}},
\bauthor{\bsnm{Yin}, \binits{J.-X.}},
\bauthor{\bsnm{Chu}, \binits{J.-H.}},
\bauthor{\bsnm{Hashimoto}, \binits{M.}},
\bauthor{\bsnm{Lu}, \binits{D.}},
\bauthor{\bsnm{Jozwiak}, \binits{C.}},
\bauthor{\bsnm{Bostwick}, \binits{A.}},
\bauthor{\bsnm{Rotenberg}, \binits{E.}},
\bauthor{\bsnm{Granroth}, \binits{G.E.}},
\bauthor{\bsnm{Yan}, \binits{B.}},
\bauthor{\bsnm{Birgeneau}, \binits{R.J.}},
\bauthor{\bsnm{Dai}, \binits{P.}},
\bauthor{\bsnm{Yi}, \binits{M.}}:
\batitle{Magnetism and charge density wave order in kagome {FeGe}}.
\bjtitle{Nature Physics}
\bvolume{19}(\bissue{6}),
\bfpage{814}--\blpage{822}
(\byear{2023})
\doiurl{10.1038/s41567-023-01985-w} .
Accessed 2024-06-26
\end{barticle}
\endbibitem

\bibitem[\protect\citeauthoryear{Chen et~al.}{2024}]{chen_competing_2024}
\begin{barticle}
\bauthor{\bsnm{Chen}, \binits{L.}},
\bauthor{\bsnm{Teng}, \binits{X.}},
\bauthor{\bsnm{Tan}, \binits{H.}},
\bauthor{\bsnm{Winn}, \binits{B.L.}},
\bauthor{\bsnm{Granroth}, \binits{G.E.}},
\bauthor{\bsnm{Ye}, \binits{F.}},
\bauthor{\bsnm{Yu}, \binits{D.H.}},
\bauthor{\bsnm{Mole}, \binits{R.A.}},
\bauthor{\bsnm{Gao}, \binits{B.}},
\bauthor{\bsnm{Yan}, \binits{B.}},
\bauthor{\bsnm{Yi}, \binits{M.}},
\bauthor{\bsnm{Dai}, \binits{P.}}:
\batitle{Competing itinerant and local spin interactions in kagome metal {FeGe}}.
\bjtitle{Nature Communications}
\bvolume{15}(\bissue{1}),
\bfpage{1918}
(\byear{2024})
\doiurl{10.1038/s41467-023-44190-2} .
Accessed 2024-06-26
\end{barticle}
\endbibitem

\bibitem[\protect\citeauthoryear{Wu et~al.}{2024}]{PhysRevX.14.011043}
\begin{barticle}
\bauthor{\bsnm{Wu}, \binits{S.}},
\bauthor{\bsnm{Klemm}, \binits{M.L.}},
\bauthor{\bsnm{Shah}, \binits{J.}},
\bauthor{\bsnm{Ritz}, \binits{E.T.}},
\bauthor{\bsnm{Duan}, \binits{C.}},
\bauthor{\bsnm{Teng}, \binits{X.}},
\bauthor{\bsnm{Gao}, \binits{B.}},
\bauthor{\bsnm{Ye}, \binits{F.}},
\bauthor{\bsnm{Matsuda}, \binits{M.}},
\bauthor{\bsnm{Li}, \binits{F.}},
\bauthor{\bsnm{Xu}, \binits{X.}},
\bauthor{\bsnm{Yi}, \binits{M.}},
\bauthor{\bsnm{Birol}, \binits{T.}},
\bauthor{\bsnm{Dai}, \binits{P.}},
\bauthor{\bsnm{Blumberg}, \binits{G.}}:
\batitle{Symmetry breaking and ascending in the magnetic kagome metal fege}.
\bjtitle{Phys. Rev. X}
\bvolume{14},
\bfpage{011043}
(\byear{2024})
\doiurl{10.1103/PhysRevX.14.011043}
\end{barticle}
\endbibitem

\bibitem[\protect\citeauthoryear{Wenzel et~al.}{2024}]{PhysRevLett.132.266505}
\begin{barticle}
\bauthor{\bsnm{Wenzel}, \binits{M.}},
\bauthor{\bsnm{Uykur}, \binits{E.}},
\bauthor{\bsnm{Tsirlin}, \binits{A.A.}},
\bauthor{\bsnm{Pal}, \binits{S.}},
\bauthor{\bsnm{Roy}, \binits{R.M.}},
\bauthor{\bsnm{Yi}, \binits{C.}},
\bauthor{\bsnm{Shekhar}, \binits{C.}},
\bauthor{\bsnm{Felser}, \binits{C.}},
\bauthor{\bsnm{Pronin}, \binits{A.V.}},
\bauthor{\bsnm{Dressel}, \binits{M.}}:
\batitle{Intriguing low-temperature phase in the antiferromagnetic kagome metal fege}.
\bjtitle{Phys. Rev. Lett.}
\bvolume{132},
\bfpage{266505}
(\byear{2024})
\doiurl{10.1103/PhysRevLett.132.266505}
\end{barticle}
\endbibitem

\bibitem[\protect\citeauthoryear{Teng et~al.}{2024}]{teng_spin-charge-lattice_2024}
\begin{barticle}
\bauthor{\bsnm{Teng}, \binits{X.}},
\bauthor{\bsnm{Tam}, \binits{D.W.}},
\bauthor{\bsnm{Chen}, \binits{L.}},
\bauthor{\bsnm{Tan}, \binits{H.}},
\bauthor{\bsnm{Xie}, \binits{Y.}},
\bauthor{\bsnm{Gao}, \binits{B.}},
\bauthor{\bsnm{Granroth}, \binits{G.E.}},
\bauthor{\bsnm{Ivanov}, \binits{A.}},
\bauthor{\bsnm{Bourges}, \binits{P.}},
\bauthor{\bsnm{Yan}, \binits{B.}},
\bauthor{\bsnm{Yi}, \binits{M.}},
\bauthor{\bsnm{Dai}, \binits{P.}}:
\batitle{Spin-charge-lattice coupling across the charge density wave transition in a kagome lattice antiferromagnet}.
\bjtitle{Phys. Rev. Lett.}
\bvolume{133},
\bfpage{046502}
(\byear{2024})
\doiurl{10.1103/PhysRevLett.133.046502}
\end{barticle}
\endbibitem

\bibitem[\protect\citeauthoryear{Lin et~al.}{2024}]{PhysRevB.110.L041121}
\begin{barticle}
\bauthor{\bsnm{Lin}, \binits{Y.-P.}},
\bauthor{\bsnm{Liu}, \binits{C.}},
\bauthor{\bsnm{Moore}, \binits{J.E.}}:
\batitle{Complex magnetic and spatial symmetry breaking from correlations in kagome flat bands}.
\bjtitle{Phys. Rev. B}
\bvolume{110},
\bfpage{041121}
(\byear{2024})
\doiurl{10.1103/PhysRevB.110.L041121}
\end{barticle}
\endbibitem

\bibitem[\protect\citeauthoryear{Tranquada}{2020}]{tranquada_cuprate_2020}
\begin{barticle}
\bauthor{\bsnm{Tranquada}, \binits{J.M.}}:
\batitle{Cuprate superconductors as viewed through a striped lens}.
\bjtitle{Advances in Physics}
\bvolume{69}(\bissue{4}),
\bfpage{437}--\blpage{509}
(\byear{2020})
\doiurl{10.1080/00018732.2021.1935698} .
\bcomment{Publisher: Taylor \& Francis \_eprint: https://doi.org/10.1080/00018732.2021.1935698}
\end{barticle}
\endbibitem

\bibitem[\protect\citeauthoryear{Tranquada et~al.}{1994}]{tranquada_simultaneous_1994}
\begin{barticle}
\bauthor{\bsnm{Tranquada}, \binits{J.M.}},
\bauthor{\bsnm{Buttrey}, \binits{D.J.}},
\bauthor{\bsnm{Sachan}, \binits{V.}},
\bauthor{\bsnm{Lorenzo}, \binits{J.E.}}:
\batitle{Simultaneous {Ordering} of {Holes} and {Spins} in {\textbackslash}{mathrmLa}\_2ni{\textbackslash}{mathrmO}\_4.125}.
\bjtitle{Phys. Rev. Lett.}
\bvolume{73}(\bissue{7}),
\bfpage{1003}--\blpage{1006}
(\byear{1994})
\doiurl{10.1103/PhysRevLett.73.1003} .
\bcomment{Publisher: American Physical Society}
\end{barticle}
\endbibitem

\bibitem[\protect\citeauthoryear{Zhang et~al.}{2020}]{zhang_intertwined_2020}
\begin{botherref}
\oauthor{\bsnm{Zhang}, \binits{J.}},
\oauthor{\bsnm{Phelan}, \binits{D.}},
\oauthor{\bsnm{Botana}, \binits{A.S.}},
\oauthor{\bsnm{Chen}, \binits{Y.-S.}},
\oauthor{\bsnm{Zheng}, \binits{H.}},
\oauthor{\bsnm{Krogstad}, \binits{M.}},
\oauthor{\bsnm{Wang}, \binits{S.G.}},
\oauthor{\bsnm{Qiu}, \binits{Y.}},
\oauthor{\bsnm{Rodriguez-Rivera}, \binits{J.A.}},
\oauthor{\bsnm{Osborn}, \binits{R.}},
\oauthor{\bsnm{Rosenkranz}, \binits{S.}},
\oauthor{\bsnm{Norman}, \binits{M.R.}},
\oauthor{\bsnm{Mitchell}, \binits{J.F.}}:
Intertwined density waves in a metallic nickelate.
NATURE COMMUNICATIONS
\textbf{11}(1)
(2020)
\doiurl{10.1038/s41467-020-19836-0}
\end{botherref}
\endbibitem

\bibitem[\protect\citeauthoryear{Meier et~al.}{2020}]{meier_flat_2020}
\begin{barticle}
\bauthor{\bsnm{Meier}, \binits{W.R.}},
\bauthor{\bsnm{Du}, \binits{M.-H.}},
\bauthor{\bsnm{Okamoto}, \binits{S.}},
\bauthor{\bsnm{Mohanta}, \binits{N.}},
\bauthor{\bsnm{May}, \binits{A.F.}},
\bauthor{\bsnm{McGuire}, \binits{M.A.}},
\bauthor{\bsnm{Bridges}, \binits{C.A.}},
\bauthor{\bsnm{Samolyuk}, \binits{G.D.}},
\bauthor{\bsnm{Sales}, \binits{B.C.}}:
\batitle{Flat bands in the {CoSn}-type compounds}.
\bjtitle{Physical Review B}
\bvolume{102}(\bissue{7}),
\bfpage{075148}
(\byear{2020})
\doiurl{10.1103/PhysRevB.102.075148} .
Accessed 2024-06-26
\end{barticle}
\endbibitem

\bibitem[\protect\citeauthoryear{Miao et~al.}{2023}]{miao_signature_2023}
\begin{barticle}
\bauthor{\bsnm{Miao}, \binits{H.}},
\bauthor{\bsnm{Zhang}, \binits{T.T.}},
\bauthor{\bsnm{Li}, \binits{H.X.}},
\bauthor{\bsnm{Fabbris}, \binits{G.}},
\bauthor{\bsnm{Said}, \binits{A.H.}},
\bauthor{\bsnm{Tartaglia}, \binits{R.}},
\bauthor{\bsnm{Yilmaz}, \binits{T.}},
\bauthor{\bsnm{Vescovo}, \binits{E.}},
\bauthor{\bsnm{Yin}, \binits{J.-X.}},
\bauthor{\bsnm{Murakami}, \binits{S.}},
\bauthor{\bsnm{Feng}, \binits{X.L.}},
\bauthor{\bsnm{Jiang}, \binits{K.}},
\bauthor{\bsnm{Wu}, \binits{X.L.}},
\bauthor{\bsnm{Wang}, \binits{A.F.}},
\bauthor{\bsnm{Okamoto}, \binits{S.}},
\bauthor{\bsnm{Wang}, \binits{Y.L.}},
\bauthor{\bsnm{Lee}, \binits{H.N.}}:
\batitle{Signature of spin-phonon coupling driven charge density wave in a kagome magnet}.
\bjtitle{Nature Communications}
\bvolume{14}(\bissue{1}),
\bfpage{6183}
(\byear{2023})
\doiurl{10.1038/s41467-023-41957-5} .
Accessed 2024-06-26
\end{barticle}
\endbibitem

\bibitem[\protect\citeauthoryear{Wu et~al.}{2024}]{wu_annealing-tunable_2024}
\begin{barticle}
\bauthor{\bsnm{Wu}, \binits{X.}},
\bauthor{\bsnm{Mi}, \binits{X.}},
\bauthor{\bsnm{Zhang}, \binits{L.}},
\bauthor{\bsnm{Wang}, \binits{C.-W.}},
\bauthor{\bsnm{Maraytta}, \binits{N.}},
\bauthor{\bsnm{Zhou}, \binits{X.}},
\bauthor{\bsnm{He}, \binits{M.}},
\bauthor{\bsnm{Merz}, \binits{M.}},
\bauthor{\bsnm{Chai}, \binits{Y.}},
\bauthor{\bsnm{Wang}, \binits{A.}}:
\batitle{Annealing-{Tunable} {Charge} {Density} {Wave} in the {Magnetic} {Kagome} {Material} {FeGe}}.
\bjtitle{Physical Review Letters}
\bvolume{132}(\bissue{25}),
\bfpage{256501}
(\byear{2024})
\doiurl{10.1103/PhysRevLett.132.256501} .
Accessed 2024-06-26
\end{barticle}
\endbibitem

\bibitem[\protect\citeauthoryear{Chen et~al.}{2024}]{chen_discovery_2024}
\begin{barticle}
\bauthor{\bsnm{Chen}, \binits{Z.}},
\bauthor{\bsnm{Wu}, \binits{X.}},
\bauthor{\bsnm{Zhou}, \binits{S.}},
\bauthor{\bsnm{Zhang}, \binits{J.}},
\bauthor{\bsnm{Yin}, \binits{R.}},
\bauthor{\bsnm{Li}, \binits{Y.}},
\bauthor{\bsnm{Li}, \binits{M.}},
\bauthor{\bsnm{Gong}, \binits{J.}},
\bauthor{\bsnm{He}, \binits{M.}},
\bauthor{\bsnm{Chai}, \binits{Y.}},
\bauthor{\bsnm{Zhou}, \binits{X.}},
\bauthor{\bsnm{Wang}, \binits{Y.}},
\bauthor{\bsnm{Wang}, \binits{A.}},
\bauthor{\bsnm{Yan}, \binits{Y.-J.}},
\bauthor{\bsnm{Feng}, \binits{D.-L.}}:
\batitle{Discovery of a long-ranged charge order with 1/4 {Ge1}-dimerization in an antiferromagnetic {Kagome} metal}.
\bjtitle{Nature Communications}
\bvolume{15},
\bfpage{6262}
(\byear{2024})
\doiurl{10.1038/s41467-024-50661-x}
\end{barticle}
\endbibitem

\bibitem[\protect\citeauthoryear{Shi et~al.}{2023}]{shi_disordered_2023}
\begin{botherref}
\oauthor{\bsnm{Shi}, \binits{C.}},
\oauthor{\bsnm{Liu}, \binits{Y.}},
\oauthor{\bsnm{Maity}, \binits{B.B.}},
\oauthor{\bsnm{Wang}, \binits{Q.}},
\oauthor{\bsnm{Kotla}, \binits{S.R.}},
\oauthor{\bsnm{Ramakrishnan}, \binits{S.}},
\oauthor{\bsnm{Eisele}, \binits{C.}},
\oauthor{\bsnm{Agarwal}, \binits{H.}},
\oauthor{\bsnm{Noohinejad}, \binits{L.}},
\oauthor{\bsnm{Tao}, \binits{Q.}},
\oauthor{\bsnm{Kang}, \binits{B.}},
\oauthor{\bsnm{Lou}, \binits{Z.}},
\oauthor{\bsnm{Yang}, \binits{X.}},
\oauthor{\bsnm{Qi}, \binits{Y.}},
\oauthor{\bsnm{Lin}, \binits{X.}},
\oauthor{\bsnm{Xu}, \binits{Z.-A.}},
\oauthor{\bsnm{Thamizhavel}, \binits{A.}},
\oauthor{\bsnm{Cao}, \binits{G.-H.}},
\oauthor{\bsnm{Smaalen}, \binits{S.}},
\oauthor{\bsnm{Cao}, \binits{S.}},
\oauthor{\bsnm{Bao}, \binits{J.-K.}}:
Disordered structure for long-range charge density wave order in annealed crystals of magnetic kagome {FeGe}.
arXiv.
arXiv:2308.09034 [cond-mat]
(2023).
\url{http://arxiv.org/abs/2308.09034}
Accessed 2024-06-26
\end{botherref}
\endbibitem

\bibitem[\protect\citeauthoryear{Smejkal et~al.}{2022}]{smejkal_anomalous_2022}
\begin{barticle}
\bauthor{\bsnm{Smejkal}, \binits{L.}},
\bauthor{\bsnm{MacDonald}, \binits{A.H.}},
\bauthor{\bsnm{Sinova}, \binits{J.}},
\bauthor{\bsnm{Nakatsuji}, \binits{S.}},
\bauthor{\bsnm{Jungwirth}, \binits{T.}}:
\batitle{Anomalous {Hall} antiferromagnets}.
\bjtitle{NATURE REVIEWS MATERIALS}
\bvolume{7}(\bissue{6}),
\bfpage{482}--\blpage{496}
(\byear{2022})
\doiurl{10.1038/s41578-022-00430-3}
\end{barticle}
\endbibitem

\bibitem[\protect\citeauthoryear{Wilson and Ortiz}{2024}]{wilson_avsb_2024}
\begin{barticle}
\bauthor{\bsnm{Wilson}, \binits{S.D.}},
\bauthor{\bsnm{Ortiz}, \binits{B.R.}}:
\batitle{{AV$_3$} {Sb$_5$} kagome superconductors}.
\bjtitle{Nature Review Materials}
\bvolume{9}(\bissue{6}),
\bfpage{420}--\blpage{432}
(\byear{2024})
\doiurl{10.1038/s41578-024-00677-y}
\end{barticle}
\endbibitem

\bibitem[\protect\citeauthoryear{Yang et~al.}{2020}]{yang_giant_2020}
\begin{barticle}
\bauthor{\bsnm{Yang}, \binits{S.-Y.}},
\bauthor{\bsnm{Wang}, \binits{Y.}},
\bauthor{\bsnm{Ortiz}, \binits{B.R.}},
\bauthor{\bsnm{Liu}, \binits{D.}},
\bauthor{\bsnm{Gayles}, \binits{J.}},
\bauthor{\bsnm{Derunova}, \binits{E.}},
\bauthor{\bsnm{Gonzalez-Hernandez}, \binits{R.}},
\bauthor{\bsnm{Šmejkal}, \binits{L.}},
\bauthor{\bsnm{Chen}, \binits{Y.}},
\bauthor{\bsnm{Parkin}, \binits{S.S.P.}},
\bauthor{\bsnm{Wilson}, \binits{S.D.}},
\bauthor{\bsnm{Toberer}, \binits{E.S.}},
\bauthor{\bsnm{McQueen}, \binits{T.}},
\bauthor{\bsnm{Ali}, \binits{M.N.}}:
\batitle{Giant, unconventional anomalous {Hall} effect in the metallic frustrated magnet candidate, {KV} $_{\textrm{3}}$ {Sb} $_{\textrm{5}}$}.
\bjtitle{Science Advances}
\bvolume{6}(\bissue{31}),
\bfpage{6003}
(\byear{2020})
\doiurl{10.1126/sciadv.abb6003} .
Accessed 2024-07-23
\end{barticle}
\endbibitem

\bibitem[\protect\citeauthoryear{Yu et~al.}{2021}]{yu_concurrence_2021}
\begin{barticle}
\bauthor{\bsnm{Yu}, \binits{F.H.}},
\bauthor{\bsnm{Wu}, \binits{T.}},
\bauthor{\bsnm{Wang}, \binits{Z.Y.}},
\bauthor{\bsnm{Lei}, \binits{B.}},
\bauthor{\bsnm{Zhuo}, \binits{W.Z.}},
\bauthor{\bsnm{Ying}, \binits{J.J.}},
\bauthor{\bsnm{Chen}, \binits{X.H.}}:
\batitle{Concurrence of anomalous {Hall} effect and charge density wave in a superconducting topological kagome metal}.
\bjtitle{Physical Review B}
\bvolume{104}(\bissue{4}),
\bfpage{041103}
(\byear{2021})
\doiurl{10.1103/PhysRevB.104.L041103} .
Accessed 2024-07-02
\end{barticle}
\endbibitem

\bibitem[\protect\citeauthoryear{Wang et~al.}{2021}]{wang_field-induced_2021}
\begin{barticle}
\bauthor{\bsnm{Wang}, \binits{Q.}},
\bauthor{\bsnm{Neubauer}, \binits{K.J.}},
\bauthor{\bsnm{Duan}, \binits{C.}},
\bauthor{\bsnm{Yin}, \binits{Q.}},
\bauthor{\bsnm{Fujitsu}, \binits{S.}},
\bauthor{\bsnm{Hosono}, \binits{H.}},
\bauthor{\bsnm{Ye}, \binits{F.}},
\bauthor{\bsnm{Zhang}, \binits{R.}},
\bauthor{\bsnm{Chi}, \binits{S.}},
\bauthor{\bsnm{Krycka}, \binits{K.}},
\bauthor{\bsnm{Lei}, \binits{H.}},
\bauthor{\bsnm{Dai}, \binits{P.}}:
\batitle{Field-induced topological {Hall} effect and double-fan spin structure with a c-axis component in the metallic kagome antiferromagnetic compound {\textbackslash}{mathrmY}{\textbackslash}{mathrmMn}\_6{\textbackslash}{mathrmSn}\_6}.
\bjtitle{Phys. Rev. B}
\bvolume{103}(\bissue{1}),
\bfpage{014416}
(\byear{2021})
\doiurl{10.1103/PhysRevB.103.014416} .
\bcomment{Publisher: American Physical Society}
\end{barticle}
\endbibitem

\bibitem[\protect\citeauthoryear{Ghimire et~al.}{2020}]{ghimire_competing_2020}
\begin{barticle}
\bauthor{\bsnm{Ghimire}, \binits{N.J.}},
\bauthor{\bsnm{Dally}, \binits{R.L.}},
\bauthor{\bsnm{Poudel}, \binits{L.}},
\bauthor{\bsnm{Jones}, \binits{D.C.}},
\bauthor{\bsnm{Michel}, \binits{D.}},
\bauthor{\bsnm{Magar}, \binits{N.T.}},
\bauthor{\bsnm{Bleuel}, \binits{M.}},
\bauthor{\bsnm{McGuire}, \binits{M.A.}},
\bauthor{\bsnm{Jiang}, \binits{J.S.}},
\bauthor{\bsnm{Mitchell}, \binits{J.F.}},
\bauthor{\bsnm{Lynn}, \binits{J.W.}},
\bauthor{\bsnm{Mazin}, \binits{I.I.}}:
\batitle{Competing magnetic phases and fluctuation-driven scalar spin chirality in the kagome metal {YMn}$_{\textrm{6}}${Sn}$_{\textrm{6}}$}.
\bjtitle{Science Advances}
\bvolume{6}(\bissue{51}),
\bfpage{2680}
(\byear{2020})
\doiurl{10.1126/sciadv.abe2680} .
\bcomment{\_eprint: https://www.science.org/doi/pdf/10.1126/sciadv.abe2680}
\end{barticle}
\endbibitem

\bibitem[\protect\citeauthoryear{Dally et~al.}{2021}]{dally_chiral_2021}
\begin{barticle}
\bauthor{\bsnm{Dally}, \binits{R.L.}},
\bauthor{\bsnm{Lynn}, \binits{J.W.}},
\bauthor{\bsnm{Ghimire}, \binits{N.J.}},
\bauthor{\bsnm{Michel}, \binits{D.}},
\bauthor{\bsnm{Siegfried}, \binits{P.}},
\bauthor{\bsnm{Mazin}, \binits{I.I.}}:
\batitle{Chiral properties of the zero-field spiral state and field-induced magnetic phases of the itinerant kagome metal {\textbackslash}{mathrmYMn}\_6{\textbackslash}{mathrmSn}\_6}.
\bjtitle{Phys. Rev. B}
\bvolume{103}(\bissue{9}),
\bfpage{094413}
(\byear{2021})
\doiurl{10.1103/PhysRevB.103.094413} .
\bcomment{Publisher: American Physical Society}
\end{barticle}
\endbibitem

\bibitem[\protect\citeauthoryear{Paddison et~al.}{2022}]{paddison_magnetic_2022}
\begin{barticle}
\bauthor{\bsnm{Paddison}, \binits{J.A.M.}},
\bauthor{\bsnm{Rai}, \binits{B.K.}},
\bauthor{\bsnm{May}, \binits{A.F.}},
\bauthor{\bsnm{Calder}, \binits{S.}},
\bauthor{\bsnm{Stone}, \binits{M.B.}},
\bauthor{\bsnm{Frontzek}, \binits{M.D.}},
\bauthor{\bsnm{Christianson}, \binits{A.D.}}:
\batitle{Magnetic {Interactions} of the {Centrosymmetric} {Skyrmion} {Material} {\textbackslash}{mathrmGd}\_2{\textbackslash}{mathrmPdSi}\_3}.
\bjtitle{Phys. Rev. Lett.}
\bvolume{129}(\bissue{13}),
\bfpage{137202}
(\byear{2022})
\doiurl{10.1103/PhysRevLett.129.137202} .
\bcomment{Publisher: American Physical Society}
\end{barticle}
\endbibitem

\bibitem[\protect\citeauthoryear{Liu et~al.}{2020}]{liu_-plane_2020}
\begin{barticle}
\bauthor{\bsnm{Liu}, \binits{P.}},
\bauthor{\bsnm{Klemm}, \binits{M.L.}},
\bauthor{\bsnm{Tian}, \binits{L.}},
\bauthor{\bsnm{Lu}, \binits{X.}},
\bauthor{\bsnm{Song}, \binits{Y.}},
\bauthor{\bsnm{Tam}, \binits{D.W.}},
\bauthor{\bsnm{Schmalzl}, \binits{K.}},
\bauthor{\bsnm{Park}, \binits{J.T.}},
\bauthor{\bsnm{Li}, \binits{Y.}},
\bauthor{\bsnm{Tan}, \binits{G.}},
\bauthor{\bsnm{Su}, \binits{Y.}},
\bauthor{\bsnm{Bourdarot}, \binits{F.}},
\bauthor{\bsnm{Zhao}, \binits{Y.}},
\bauthor{\bsnm{Lynn}, \binits{J.W.}},
\bauthor{\bsnm{Birgeneau}, \binits{R.J.}},
\bauthor{\bsnm{Dai}, \binits{P.}}:
\batitle{In-plane uniaxial pressure-induced out-of-plane antiferromagnetic moment and critical fluctuations in {BaFe2As2}}.
\bjtitle{Nature Communications}
\bvolume{11}(\bissue{1}),
\bfpage{5728}
(\byear{2020})
\doiurl{10.1038/s41467-020-19421-5} .
Accessed 2024-07-01
\end{barticle}
\endbibitem

\bibitem[\protect\citeauthoryear{Lipscombe et~al.}{2010}]{lipscombe_anisotropic_2010}
\begin{barticle}
\bauthor{\bsnm{Lipscombe}, \binits{O.J.}},
\bauthor{\bsnm{Harriger}, \binits{L.W.}},
\bauthor{\bsnm{Freeman}, \binits{P.G.}},
\bauthor{\bsnm{Enderle}, \binits{M.}},
\bauthor{\bsnm{Zhang}, \binits{C.}},
\bauthor{\bsnm{Wang}, \binits{M.}},
\bauthor{\bsnm{Egami}, \binits{T.}},
\bauthor{\bsnm{Hu}, \binits{J.}},
\bauthor{\bsnm{Xiang}, \binits{T.}},
\bauthor{\bsnm{Norman}, \binits{M.R.}},
\bauthor{\bsnm{Dai}, \binits{P.}}:
\batitle{Anisotropic neutron spin resonance in superconducting {BaFe} 1.9 {Ni} 0.1 {As} 2}.
\bjtitle{Physical Review B}
\bvolume{82}(\bissue{6}),
\bfpage{064515}
(\byear{2010})
\doiurl{10.1103/PhysRevB.82.064515} .
Accessed 2024-07-01
\end{barticle}
\endbibitem

\bibitem[\protect\citeauthoryear{Berghezan et~al.}{1961}]{berghezan_transmission_1961}
\begin{barticle}
\bauthor{\bsnm{Berghezan}, \binits{A.}},
\bauthor{\bsnm{Fourdeux}, \binits{A.}},
\bauthor{\bsnm{Amelinckx}, \binits{S.}}:
\batitle{Transmission electron microscopy studies of dislocations and stacking faults in a hexagonal metal: {Zinc}}.
\bjtitle{Acta Metallurgica}
\bvolume{9}(\bissue{5}),
\bfpage{464}--\blpage{490}
(\byear{1961})
\doiurl{10.1016/0001-6160(61)90142-0}
\end{barticle}
\endbibitem

\bibitem[\protect\citeauthoryear{Yin et~al.}{2017}]{yin_comprehensive_2017}
\begin{barticle}
\bauthor{\bsnm{Yin}, \binits{B.}},
\bauthor{\bsnm{Wu}, \binits{Z.}},
\bauthor{\bsnm{Curtin}, \binits{W.A.}}:
\batitle{Comprehensive first-principles study of stable stacking faults in hcp metals}.
\bjtitle{Acta Materialia}
\bvolume{123},
\bfpage{223}--\blpage{234}
(\byear{2017})
\doiurl{10.1016/j.actamat.2016.10.042}
\end{barticle}
\endbibitem

\bibitem[\protect\citeauthoryear{Howie}{1979}]{howie_image_1979}
\begin{barticle}
\bauthor{\bsnm{Howie}, \binits{A.}}:
\batitle{Image {Contrast} {And} {Localized} {Signal} {Selection} {Techniques}}.
\bjtitle{Journal of Microscopy}
\bvolume{117}(\bissue{1}),
\bfpage{11}--\blpage{23}
(\byear{1979})
\doiurl{10.1111/j.1365-2818.1979.tb00228.x} .
\bcomment{\_eprint: https://onlinelibrary.wiley.com/doi/pdf/10.1111/j.1365-2818.1979.tb00228.x}
\end{barticle}
\endbibitem

\bibitem[\protect\citeauthoryear{Hartel et~al.}{1996}]{hartel_conditions_1996}
\begin{barticle}
\bauthor{\bsnm{Hartel}, \binits{P.}},
\bauthor{\bsnm{Rose}, \binits{H.}},
\bauthor{\bsnm{Dinges}, \binits{C.}}:
\batitle{Conditions and reasons for incoherent imaging in {STEM}}.
\bjtitle{Ultramicroscopy}
\bvolume{63}(\bissue{2}),
\bfpage{93}--\blpage{114}
(\byear{1996})
\doiurl{10.1016/0304-3991(96)00020-4}
\end{barticle}
\endbibitem

\bibitem[\protect\citeauthoryear{Cahn}{1965}]{cahn_direct_1965}
\begin{botherref}
\oauthor{\bsnm{Cahn}, \binits{R.W.}}:
The direct observation of dislocations: {S}. {Amelinckx}, ({Academic} press, 1964) x + 487 pages. \$17.00.
Journal of Nuclear Materials
\textbf{17}
(1965)
\end{botherref}
\endbibitem

\bibitem[\protect\citeauthoryear{Hart et~al.}{2024}]{hart_real-space_2024}
\begin{barticle}
\bauthor{\bsnm{Hart}, \binits{J.L.}},
\bauthor{\bsnm{Pan}, \binits{H.}},
\bauthor{\bsnm{Siddique}, \binits{S.}},
\bauthor{\bsnm{Schnitzer}, \binits{N.}},
\bauthor{\bsnm{Mallayya}, \binits{K.}},
\bauthor{\bsnm{Xu}, \binits{S.}},
\bauthor{\bsnm{Kourkoutis}, \binits{L.F.}},
\bauthor{\bsnm{Kim}, \binits{E.-a.}},
\bauthor{\bsnm{Cha}, \binits{J.J.}}:
\batitle{Real-space visualization of a defect-mediated charge density wave transition}.
\bjtitle{Proceedings of the National Academy of Sciences}
\bvolume{121}(\bissue{33}),
\bfpage{2402129121}
(\byear{2024})
\doiurl{10.1073/pnas.2402129121} .
\bcomment{\_eprint: https://www.pnas.org/doi/pdf/10.1073/pnas.2402129121}
\end{barticle}
\endbibitem

\bibitem[\protect\citeauthoryear{Suter and Wojek}{2012}]{suter_musrfit_2012}
\begin{barticle}
\bauthor{\bsnm{Suter}, \binits{A.}},
\bauthor{\bsnm{Wojek}, \binits{B.M.}}:
\batitle{Musrfit: {A} {Free} {Platform}-{Independent} {Framework} for {$\mu$SR} {Data} {Analysis}}.
\bjtitle{Physics Procedia}
\bvolume{30},
\bfpage{69}--\blpage{73}
(\byear{2012})
\doiurl{10.1016/j.phpro.2012.04.042} .
Accessed 2024-07-01
\end{barticle}
\endbibitem

\bibitem[\protect\citeauthoryear{Wang}{2023}]{wang_enhanced_2023}
\begin{barticle}
\bauthor{\bsnm{Wang}, \binits{Y.}}:
\batitle{Enhanced spin-polarization via partial {Ge}-dimerization as the driving force of the charge density wave in {FeGe}}.
\bjtitle{Physical Review Materials}
\bvolume{7}(\bissue{10}),
\bfpage{104006}
(\byear{2023})
\doiurl{10.1103/PhysRevMaterials.7.104006} .
Accessed 2024-06-26
\end{barticle}
\endbibitem

\bibitem[\protect\citeauthoryear{Zhou et~al.}{2023}]{zhou_magnetic_2023}
\begin{barticle}
\bauthor{\bsnm{Zhou}, \binits{H.}},
\bauthor{\bsnm{Yan}, \binits{S.}},
\bauthor{\bsnm{Fan}, \binits{D.}},
\bauthor{\bsnm{Wang}, \binits{D.}},
\bauthor{\bsnm{Wan}, \binits{X.}}:
\batitle{Magnetic interactions and possible structural distortion in kagome {FeGe} from first-principles calculations and symmetry analysis}.
\bjtitle{Phys. Rev. B}
\bvolume{108}(\bissue{3}),
\bfpage{035138}
(\byear{2023})
\doiurl{10.1103/PhysRevB.108.035138} .
\bcomment{Publisher: American Physical Society}
\end{barticle}
\endbibitem

\bibitem[\protect\citeauthoryear{Yi et~al.}{2024}]{yi_polarized_2024}
\begin{botherref}
\oauthor{\bsnm{Yi}, \binits{S.}},
\oauthor{\bsnm{Liao}, \binits{Z.}},
\oauthor{\bsnm{Wang}, \binits{Q.}},
\oauthor{\bsnm{Ma}, \binits{H.}},
\oauthor{\bsnm{Liu}, \binits{J.}},
\oauthor{\bsnm{Teng}, \binits{X.}},
\oauthor{\bsnm{Dai}, \binits{P.}},
\oauthor{\bsnm{Dai}, \binits{Y.}},
\oauthor{\bsnm{Zhao}, \binits{J.}},
\oauthor{\bsnm{Qi}, \binits{Y.}},
\oauthor{\bsnm{Xu}, \binits{B.}},
\oauthor{\bsnm{Qiu}, \binits{X.}}:
Polarized {Charge} {Dynamics} of a {Novel} {Charge} {Density} {Wave} in {Kagome} {FeGe}.
arXiv.
arXiv:2403.09950 [cond-mat]
(2024).
\url{http://arxiv.org/abs/2403.09950}
Accessed 2024-06-26
\end{botherref}
\endbibitem

\bibitem[\protect\citeauthoryear{Zhao et~al.}{2023}]{zhao_photoemission_2023}
\begin{botherref}
\oauthor{\bsnm{Zhao}, \binits{Z.}},
\oauthor{\bsnm{Li}, \binits{T.}},
\oauthor{\bsnm{Li}, \binits{P.}},
\oauthor{\bsnm{Wu}, \binits{X.}},
\oauthor{\bsnm{Yao}, \binits{J.}},
\oauthor{\bsnm{Chen}, \binits{Z.}},
\oauthor{\bsnm{Cui}, \binits{S.}},
\oauthor{\bsnm{Sun}, \binits{Z.}},
\oauthor{\bsnm{Yang}, \binits{Y.}},
\oauthor{\bsnm{Jiang}, \binits{Z.}},
\oauthor{\bsnm{Liu}, \binits{Z.}},
\oauthor{\bsnm{Louat}, \binits{A.}},
\oauthor{\bsnm{Kim}, \binits{T.}},
\oauthor{\bsnm{Cacho}, \binits{C.}},
\oauthor{\bsnm{Wang}, \binits{A.}},
\oauthor{\bsnm{Wang}, \binits{Y.}},
\oauthor{\bsnm{Shen}, \binits{D.}},
\oauthor{\bsnm{Jiang}, \binits{J.}},
\oauthor{\bsnm{Feng}, \binits{D.}}:
Photoemission {Evidence} of a {Novel} {Charge} {Order} in {Kagome} {Metal} {FeGe}.
arXiv.
Version Number: 1
(2023).
\doiurl{10.48550/ARXIV.2308.08336} .
\url{https://arxiv.org/abs/2308.08336}
Accessed 2024-06-26
\end{botherref}
\endbibitem

\bibitem[\protect\citeauthoryear{Feng et~al.}{2021}]{feng_chiral_2021}
\begin{barticle}
\bauthor{\bsnm{Feng}, \binits{X.}},
\bauthor{\bsnm{Jiang}, \binits{K.}},
\bauthor{\bsnm{Wang}, \binits{Z.}},
\bauthor{\bsnm{Hu}, \binits{J.}}:
\batitle{Chiral flux phase in the {Kagome} superconductor {AV3Sb5}}.
\bjtitle{Science Bulletin}
\bvolume{66}(\bissue{14}),
\bfpage{1384}--\blpage{1388}
(\byear{2021})
\doiurl{10.1016/j.scib.2021.04.043} .
Accessed 2024-07-15
\end{barticle}
\endbibitem

\bibitem[\protect\citeauthoryear{Subires et~al.}{2024}]{subires_frustrated_2024}
\begin{botherref}
\oauthor{\bsnm{Subires}, \binits{D.}},
\oauthor{\bsnm{Kar}, \binits{A.}},
\oauthor{\bsnm{Korshunov}, \binits{A.}},
\oauthor{\bsnm{Fuller}, \binits{C.A.}},
\oauthor{\bsnm{Jiang}, \binits{Y.}},
\oauthor{\bsnm{Hu}, \binits{H.}},
\oauthor{\bsnm{Călugăru}, \binits{D.}},
\oauthor{\bsnm{McMonagle}, \binits{C.}},
\oauthor{\bsnm{Yi}, \binits{C.}},
\oauthor{\bsnm{Roychowdhury}, \binits{S.}},
\oauthor{\bsnm{Shekhar}, \binits{C.}},
\oauthor{\bsnm{Strempfer}, \binits{J.}},
\oauthor{\bsnm{Jana}, \binits{A.}},
\oauthor{\bsnm{Vobornik}, \binits{I.}},
\oauthor{\bsnm{Dai}, \binits{J.}},
\oauthor{\bsnm{Tallarida}, \binits{M.}},
\oauthor{\bsnm{Chernyshov}, \binits{D.}},
\oauthor{\bsnm{Bosak}, \binits{A.}},
\oauthor{\bsnm{Felser}, \binits{C.}},
\oauthor{\bsnm{Bernevig}, \binits{B.A.}},
\oauthor{\bsnm{Blanco-Canosa}, \binits{S.}}:
Frustrated charge density wave and quasi-long-range bond-orientational order in the magnetic kagome {FeGe}.
\_eprint: 2408.04452
(2024).
\url{https://arxiv.org/abs/2408.04452}
\end{botherref}
\endbibitem

\bibitem[\protect\citeauthoryear{Ye et~al.}{2018}]{ye_implementation_2018}
\begin{barticle}
\bauthor{\bsnm{Ye}, \binits{F.}},
\bauthor{\bsnm{Liu}, \binits{Y.}},
\bauthor{\bsnm{Whitfield}, \binits{R.}},
\bauthor{\bsnm{Osborn}, \binits{R.}},
\bauthor{\bsnm{Rosenkranz}, \binits{S.}}:
\batitle{Implementation of cross correlation for energy discrimination on the time-of-flight spectrometer {CORELLI}}.
\bjtitle{Journal of Applied Crystallography}
\bvolume{51}(\bissue{2}),
\bfpage{315}--\blpage{322}
(\byear{2018})
\doiurl{10.1107/S160057671800403X} .
Accessed 2024-07-02
\end{barticle}
\endbibitem

\bibitem[\protect\citeauthoryear{Cao et~al.}{2019}]{cryst9010005}
\begin{botherref}
\oauthor{\bsnm{Cao}, \binits{H.}},
\oauthor{\bsnm{Chakoumakos}, \binits{B.C.}},
\oauthor{\bsnm{Andrews}, \binits{K.M.}},
\oauthor{\bsnm{Wu}, \binits{Y.}},
\oauthor{\bsnm{Riedel}, \binits{R.A.}},
\oauthor{\bsnm{Hodges}, \binits{J.}},
\oauthor{\bsnm{Zhou}, \binits{W.}},
\oauthor{\bsnm{Gregory}, \binits{R.}},
\oauthor{\bsnm{Haberl}, \binits{B.}},
\oauthor{\bsnm{Molaison}, \binits{J.}},
\oauthor{\bsnm{Lynn}, \binits{G.W.}}:
Demand, a dimensional extreme magnetic neutron diffractometer at the high flux isotope reactor.
Crystals
\textbf{9}(1)
(2019)
\doiurl{10.3390/cryst9010005}
\end{botherref}
\endbibitem

\bibitem[\protect\citeauthoryear{Boothroyd}{2020}]{boothroyd_principles_2020}
\begin{bbook}
\bauthor{\bsnm{Boothroyd}, \binits{A.T.}}:
\bbtitle{Principles of Neutron Scattering from Condensed Matter},
\bedition{First edition published} edn.
\bpublisher{Oxford University Press},
\blocation{Oxford}
(\byear{2020})
\end{bbook}
\endbibitem

\bibitem[\protect\citeauthoryear{Savitzky et~al.}{2018}]{savitzky_image_2018}
\begin{barticle}
\bauthor{\bsnm{Savitzky}, \binits{B.H.}},
\bauthor{\bsnm{El~Baggari}, \binits{I.}},
\bauthor{\bsnm{Clement}, \binits{C.B.}},
\bauthor{\bsnm{Waite}, \binits{E.}},
\bauthor{\bsnm{Goodge}, \binits{B.H.}},
\bauthor{\bsnm{Baek}, \binits{D.J.}},
\bauthor{\bsnm{Sheckelton}, \binits{J.P.}},
\bauthor{\bsnm{Pasco}, \binits{C.}},
\bauthor{\bsnm{Nair}, \binits{H.}},
\bauthor{\bsnm{Schreiber}, \binits{N.J.}},
\bauthor{\bsnm{Hoffman}, \binits{J.}},
\bauthor{\bsnm{Admasu}, \binits{A.S.}},
\bauthor{\bsnm{Kim}, \binits{J.}},
\bauthor{\bsnm{Cheong}, \binits{S.-W.}},
\bauthor{\bsnm{Bhattacharya}, \binits{A.}},
\bauthor{\bsnm{Schlom}, \binits{D.G.}},
\bauthor{\bsnm{McQueen}, \binits{T.M.}},
\bauthor{\bsnm{Hovden}, \binits{R.}},
\bauthor{\bsnm{Kourkoutis}, \binits{L.F.}}:
\batitle{Image registration of low signal-to-noise cryo-{STEM} data}.
\bjtitle{Ultramicroscopy}
\bvolume{191},
\bfpage{56}--\blpage{65}
(\byear{2018})
\doiurl{10.1016/j.ultramic.2018.04.008} .
Accessed 2024-08-19
\end{barticle}
\endbibitem

\bibitem[\protect\citeauthoryear{Funni et~al.}{2021}]{funni_theory_2021}
\begin{barticle}
\bauthor{\bsnm{Funni}, \binits{S.D.}},
\bauthor{\bsnm{Yang}, \binits{Z.J.}},
\bauthor{\bsnm{Cabral}, \binits{M.J.}},
\bauthor{\bsnm{Ophus}, \binits{C.}},
\bauthor{\bsnm{Chen}, \binits{X.M.}},
\bauthor{\bsnm{Dickey}, \binits{E.C.}}:
\batitle{Theory and application of the vector pair correlation function for real-space crystallographic analysis of order/disorder correlations from {STEM} images}.
\bjtitle{APL Materials}
\bvolume{9}(\bissue{9}),
\bfpage{091110}
(\byear{2021})
\doiurl{10.1063/5.0058928} .
Accessed 2024-08-19
\end{barticle}
\endbibitem

\bibitem[\protect\citeauthoryear{Madsen and Susi}{2021}]{madsen_abtem_2021}
\begin{barticle}
\bauthor{\bsnm{Madsen}, \binits{J.}},
\bauthor{\bsnm{Susi}, \binits{T.}}:
\batitle{The {abTEM} code: transmission electron microscopy from first principles}.
\bjtitle{Open Research Europe}
\bvolume{1},
\bfpage{24}
(\byear{2021})
\doiurl{10.12688/openreseurope.13015.1} .
Accessed 2024-08-19
\end{barticle}
\endbibitem

\bibitem[\protect\citeauthoryear{Amato et~al.}{2017}]{amato_new_2017}
\begin{barticle}
\bauthor{\bsnm{Amato}, \binits{A.}},
\bauthor{\bsnm{Luetkens}, \binits{H.}},
\bauthor{\bsnm{Sedlak}, \binits{K.}},
\bauthor{\bsnm{Stoykov}, \binits{A.}},
\bauthor{\bsnm{Scheuermann}, \binits{R.}},
\bauthor{\bsnm{Elender}, \binits{M.}},
\bauthor{\bsnm{Raselli}, \binits{A.}},
\bauthor{\bsnm{Graf}, \binits{D.}}:
\batitle{The new versatile general purpose surface-muon instrument ({GPS}) based on silicon photomultipliers for \textit{$\mu$} {SR} measurements on a continuous-wave beam}.
\bjtitle{Review of Scientific Instruments}
\bvolume{88}(\bissue{9}),
\bfpage{093301}
(\byear{2017})
\doiurl{10.1063/1.4986045} .
Accessed 2024-07-01
\end{barticle}
\endbibitem

\end{thebibliography}

\begin{figure}[h]
\centering
\includegraphics[width=1\textwidth]{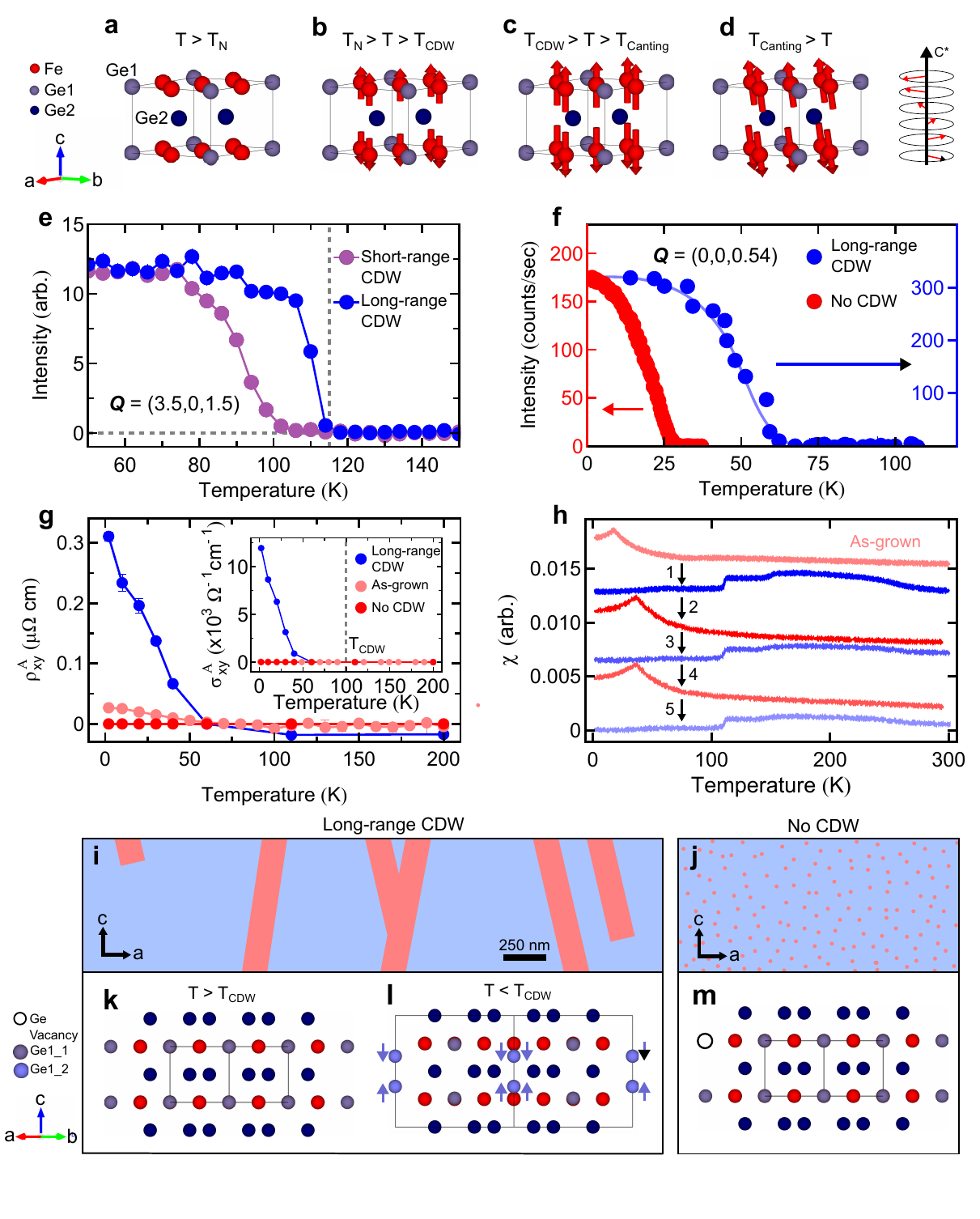}
\caption{\textbf{a-d.} Magnetic structure of FeGe in four phases. \textbf{e.} Order parameter scans of CDW Bragg peaks for two different annealing conditions. \textbf{f.} Order parameter scans of canted AFM Bragg peaks for a long-range CDW and no CDW annealed samples. \textbf{g.} Anomalous Hall resistivity as a function of temperature for as-grown sample from \cite{teng_discovery_2022}, long-range and no-CDW samples. The inset shows the anomalous Hall conductivity over the same temperature range. Anomalous Hall resistivity and conductivity are related by $\sigma^{A}_{xy}=\frac{\rho^{A}_{xy}}{(\rho^{A}_{xy})^2 + (\rho_{xx})^2}$. \textbf{h.} Magnetic susceptibility of the same sample through multiple annealing cycles. \textbf{i.} and \textbf{j.} are schematic representations of defect regions in long-range CDW samples and samples with no-CDW,  respectively. The red regions in \textbf{(i)} represent extended defects, while those in \textbf{(j)} represent uniformly distributed Ge vacancies. The blue regions in both \textbf{(i,j)} represent areas of the sample with stoichiometric FeGe. \textbf{k.} and \textbf{l.} show the refined structure for long-range ordered FeGe above and below $T_{CDW}$ respectively. \textbf{m.} Refined structure of FeGe with no long-ranged order with Ge vacancy preventing dimerization.}\label{fig1}
\end{figure}

\begin{figure}[h]
\centering
\includegraphics[width=1\textwidth]{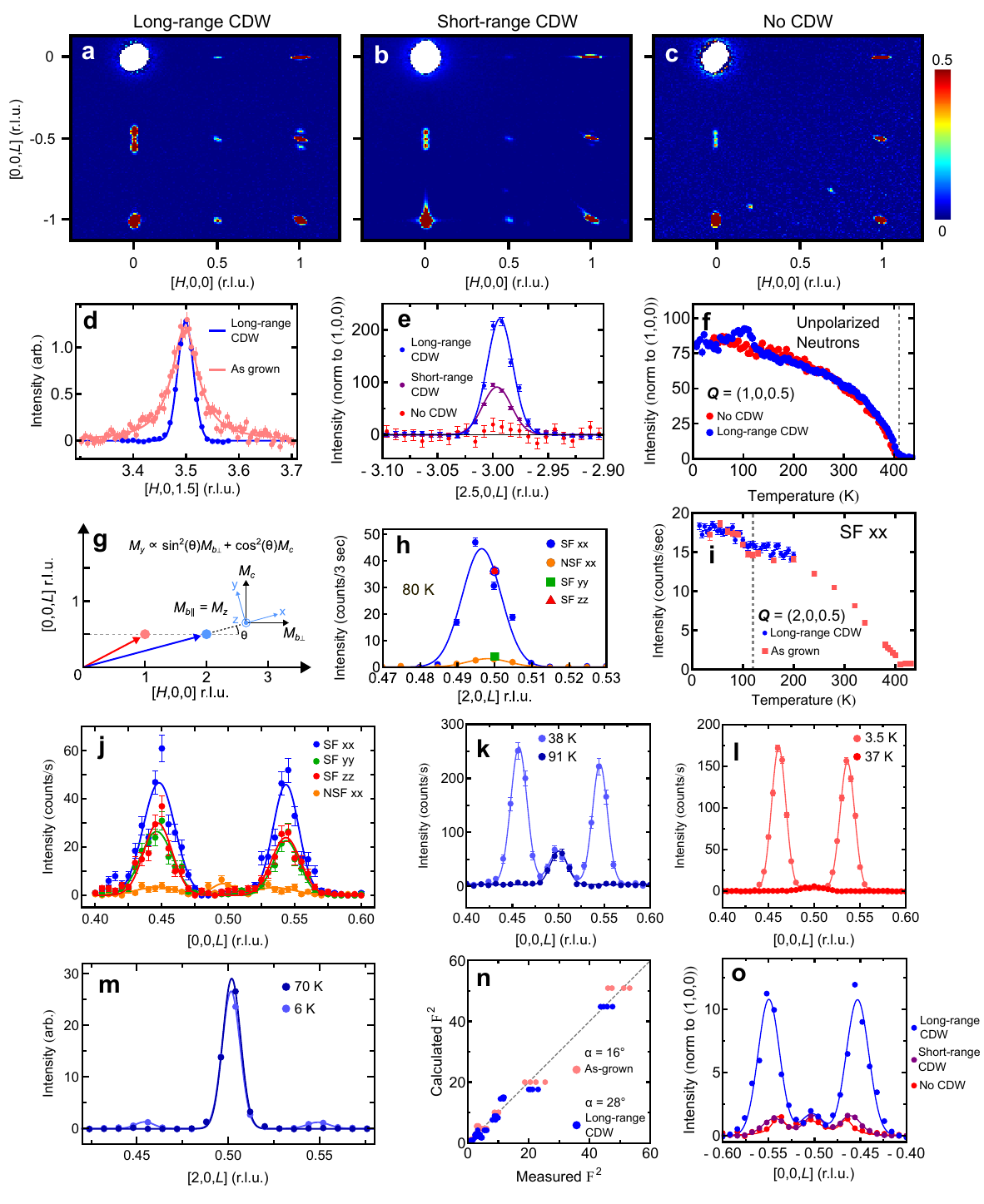}
\caption{\textbf{a-c.} 2D reciprocal space maps from neutron scattering with three samples under different annealing conditions: long-range CDW, short-range CDW and no-CDW respectively. \textbf{d}. $\pmb{Q}$-scan of CDW lineshape change between as-grown and long-range CDW samples demonstrating the long-range CDW sample is resolution limited. As-grown sample is from \cite{teng_discovery_2022}. \textbf{e.} Comparison of CDW peaks under different annealing conditions via neutron scattering. The CDW Bragg peaks are normalized to the (1,0,0) nuclear Bragg peak in each sample. \textbf{f.} Order parameter scans of two annealed samples using unpolarized neutrons normalized to the (1,0,0) nuclear Bragg peak in each sample. \textbf{g.} Schematic for polarized neutron scattering geometry within the [$H$,0,$L$] scattering plane. \textbf{h.} Polarized neutron $\pmb{Q}$-scan of an AFM Bragg peak with all polarization channels.  \textbf{i.} Order parameter scans of an as-grown sample and long-range CDW sample using polarized neutrons. \textbf{j.} Polarized $\pmb{Q}$-scans at base temperature of the incommensurate magnetic peaks along [0,0,$L$]. \textbf{k.} and \textbf{l.} show $\pmb{Q}$-scans of the incommensurate AFM Bragg peaks above and below $T_{canting}$ for long-range CDW sample and no-CDW sample respectively. \textbf{m.} Comparison of the commensurate AFM peak intensity at base temperature and just above the canted AFM transition. \textbf{n.} shows the refinement considering double-cone AFM structure for long-range CDW and as-grown samples. 
 \textbf{o.} compares incommensurate AFM peaks 
along the [0,0,$L$] direction for long-range, short-range, and no-CDW samples. 
}\label{fig2}
\end{figure}

\begin{figure}[h]
\centering
\includegraphics[width=1\textwidth]{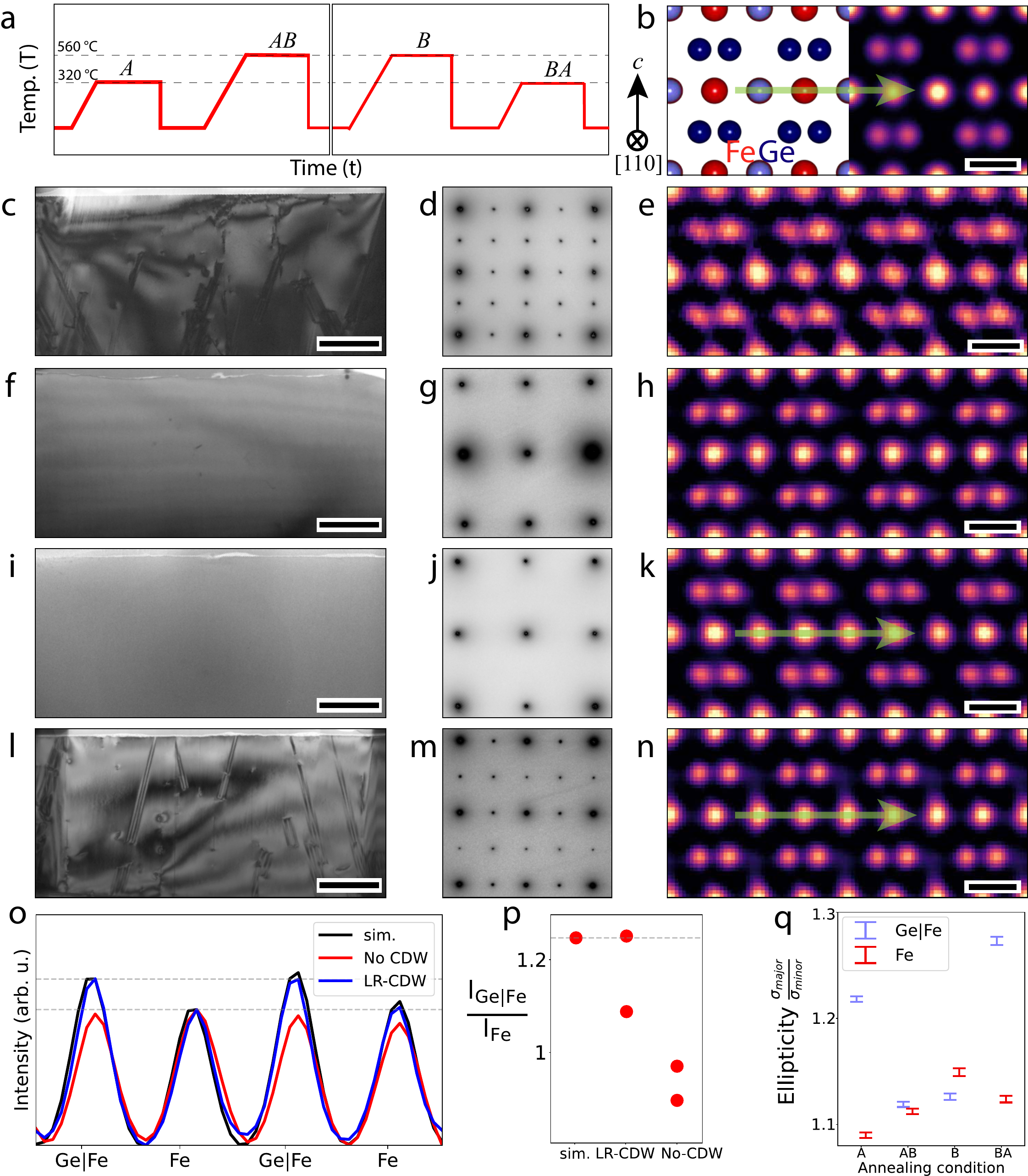}
\caption{\textbf{a.} Heat-treatment plot of the four FeGe samples used in the (S)TEM characterizations. \textbf{b.} Schematic and simulated HAADF-STEM image of FeGe in the [110] zone axis. All experimental (S)TEM images in this figure are in the [110] zone axis. BF-TEM, SAED and averaged HAADF-STEM image for different annealing conditions: \textbf{(c, d, e)} 320\textdegree C (sample $A$), \textbf{(f, g, h)} 320\textdegree C $\rightarrow$ 560\textdegree C (sample $AB$), \textbf{(i, j, k)} 560\textdegree C (sample $B$), and \textbf{(l, m, n)} 560\textdegree C $\rightarrow$ 320\textdegree C (sample $BA$). \textbf{o.} Intensity line profile from \textbf{(b, k, n)} showing variation in Ge$|$Fe and Fe columns. Only the sample annealed at 320\textdegree C (long-range (LR) CDW) shows the expected intensity profile. \textbf{p.} Ratio of intensities for Ge$|$Fe and Fe columns. Samples where the final anneal was 320\textdegree C, show the expected intensity ratio. \textbf{q.} Measured ellipticities of Fe and Ge$|$Fe columns from full HAADF-STEM images. The plot shows the standard error in our measurements. For samples with LR CDW (samples $A, BA$), the Ge$|$Fe column is more elliptical suggesting displacement of the Ge atoms in the Ge1 sites. Scale bars for \textbf{(c, f, i, l)} = 1 $\mu$m. Scale bars for \textbf{(b, e, h, k, n)} = 2 \r A. 
}\label{fig3}
\end{figure}

\begin{figure}[h]
\centering
\includegraphics[width=1\textwidth]{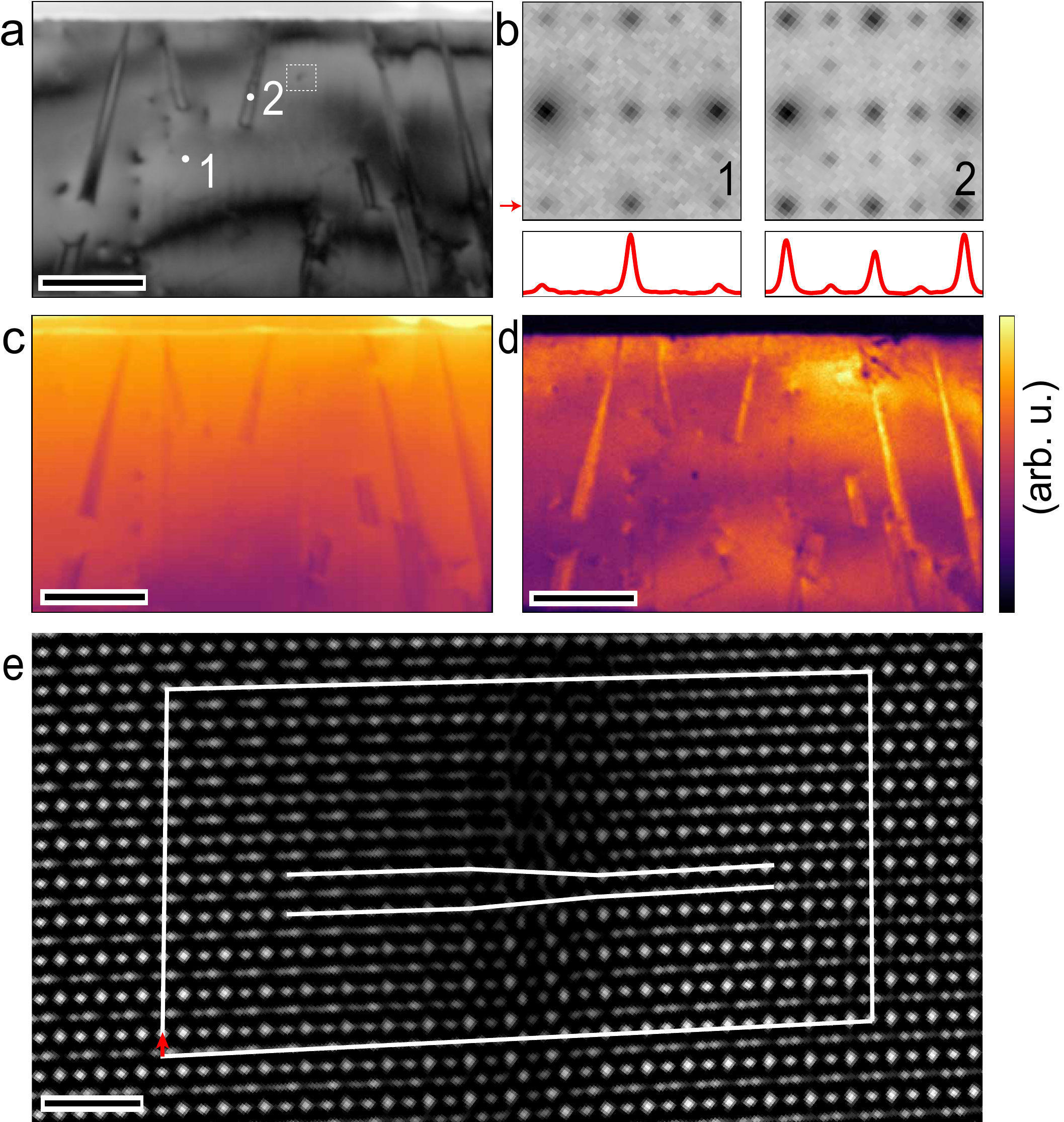}
\caption{\textbf{a.} Virtual BF-STEM image reconstructed from 4D-STEM scan of sample shown in Fig. 3l. \textbf{b.} Diffraction patterns from two probe positions. The CDW peaks are stronger in the defective region, as can be seen in the line profiles. \textbf{c.} Real-space map generated by selecting Bragg peaks in diffraction pattern. \textbf{d.} Real-space CDW map. \textbf{e.} Fourier-filtered HAADF-STEM image showing a Burger’s circuit around the dislocation in the sample boxed in \textbf{a}. The dislocation core is also drawn. Scale bars for \textbf{(a, c, d)} = 1 $\mu$m, and for \textbf{e} = 1 nm. 
}\label{fig4}
\end{figure}

\begin{figure}[h]
\centering
\includegraphics[width=1\textwidth]{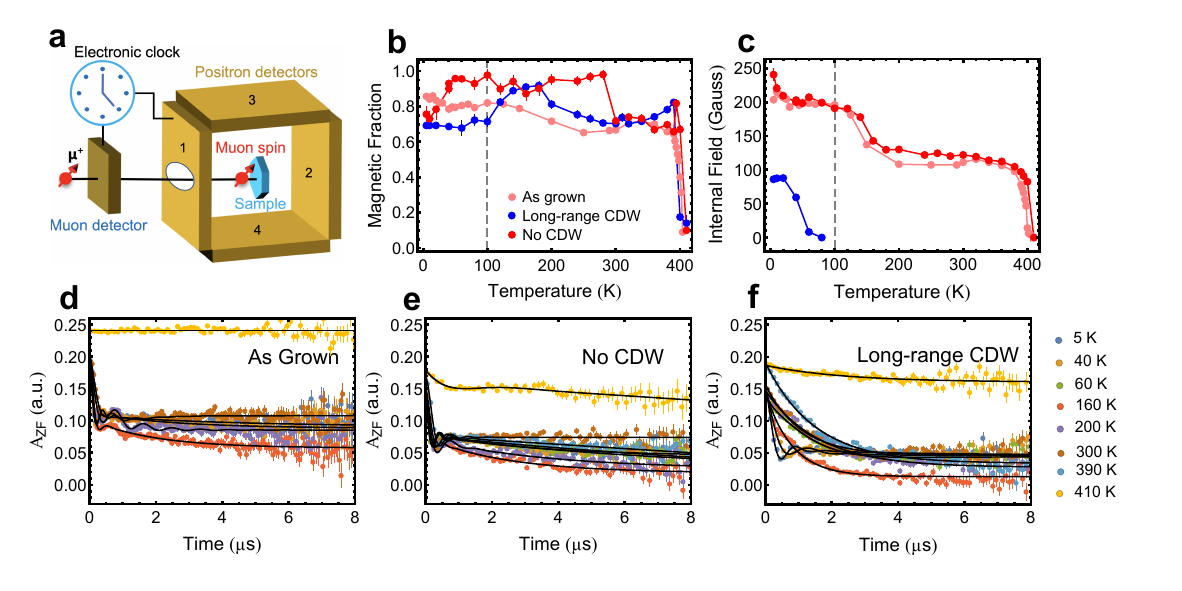}
\caption{\textbf{a.} Schematic of the $\mu$SR experimental setup. Temperature dependence of the  magnetic fractions \textbf{b.} and internal magnetic fields \textbf{c.} for three annealing conditions of FeGe. \textbf{d-f.} Zero-field (ZF) $\mu$SR time spectra for three annealing conditions of FeGe from detectors 3 and 4 at various temperatures. The vertical error bars in \textbf{(d-f)} are statistical errors of one standard deviation.}\label{fig5}
\end{figure}

\end{document}